%% file: report.tex
\documentclass[]{llncs}

\usepackage{pslatex}
\usepackage{amsmath}
\usepackage{leftidx}
\usepackage{epsfig}
\usepackage{paralist}
\usepackage{graphics}
\usepackage{stmaryrd}
\usepackage{txfonts}
\usepackage{framed}
\usepackage{makecell}
\usepackage{subfig}
\usepackage{wrapfig}
\usepackage[]{hyperref}

\usepackage[final]{commenting}

\declareauthor{ri}{Radu}{blue}
\declareauthor{mb}{Marius}{red}
\declareauthor{js}{Joseph}{magenta}

\input commands

\usepackage{setspace}

\newif\ifLongVersion\LongVersiontrue

\begin{document}

\setlength{\belowdisplayskip}{2pt} \setlength{\belowdisplayshortskip}{1pt}
\setlength{\abovedisplayskip}{2pt} \setlength{\abovedisplayshortskip}{1pt}

\title{Structural Invariants for Parametric Verification of Systems
  with Almost Linear Architectures}

\author{Marius Bozga, Radu Iosif and Joseph Sifakis}
\institute{Univ. Grenoble Alpes, CNRS, %
  Grenoble INP\footnote{Institute of Engineering Univ. Grenoble Alpes}, VERIMAG, 38000 
   Grenoble France 
}

\maketitle
\input abstract
\input body

\bibliographystyle{splncs03} \bibliography{refs}

\end{document}

%% file: commands.tex
\renewcommand{\mod}{~\mathrm{mod}~}

\newcommand{\bigO}{\mathcal{O}}

\newcommand{\rbr}{{\bf ]\!]}}
\newcommand{\lbr}{{\bf [\![}}
\newcommand{\sem}[1]{\lbr #1 \rbr}

\newcommand{\set}[1]{\{ #1 \}}
\newcommand{\tuple}[1]{\langle #1 \rangle}
\renewcommand{\vec}[1]{\mathbf #1}

\newcommand{\isdef}{\stackrel{\scriptscriptstyle{\mathsf{def}}}{=}}

\newcommand{\lang}[1]{\mathcal{L}({#1})}
\newcommand{\minlang}[1]{\mathcal{L}^{\mathrm{min}}({#1})}
\newcommand{\arrow}[2]{\xrightarrow{{\scriptscriptstyle #1}}_{{\scriptscriptstyle #2}}}

\newcommand{\nat}{{\bf \mathbb{N}}}

\newcommand{\proj}[2]{{#1}\!\!\downarrow_{{#2}}}

\renewcommand{\paragraph}[1]{\noindent{\bf #1}}

\newcommand{\I}{\mathcal{I}}

\newcommand{\true}{\top}

\newcommand{\vars}{\mathsf{Var}}
\newcommand{\Vars}{\mathsf{SVar}}

\newcommand{\preds}{\mathsf{Pred}}
\newcommand{\const}{\mathsf{Const}}

\newcommand{\pos}[1]{\left({#1}\right)^+}
\newcommand{\ppos}[1]{\left({#1}\right)^\oplus}

\newcommand{\anet}{\mathsf{N}}
\newcommand{\amarkednet}{\mathcal{N}}
\newcommand{\places}{S}
\newcommand{\trans}{T}
\newcommand{\atrans}{\mathfrak{t}}
\newcommand{\edges}{E}
\newcommand{\pre}[1]{\leftidx{^\bullet}{\text{${#1}$}}}
\newcommand{\post}[1]{{#1}^\bullet}
\newcommand{\amark}{\mathrm{m}}
\newcommand{\reach}[1]{\mathcal{R}({#1})}

\newcommand{\alltrap}[1]{\mathit{TrapInv}({#1})}

\newcommand{\init}[1]{\mathit{Init}({#1})}

\newcommand{\imt}[1]{\mathrm{Imt}({#1})}

\newcommand{\acomptype}{\mathcal{C}}

\newcommand{\typeno}[2]{{#1}^{\scriptscriptstyle{{#2}}}}
\newcommand{\typeof}[1]{\mathit{type}({#1})}
\newcommand{\ports}{\mathsf{P}}
\newcommand{\states}{\mathsf{S}}
\newcommand{\initstate}{{s_0}}
\newcommand{\rules}{\Delta}
\newcommand{\asys}{\mathcal{S}}
\newcommand{\interform}{\Gamma}
\newcommand{\viewform}[2]{\mathrm{V}_{#1}^{#2}}
\newcommand{\reachform}[2]{\Phi_{#1}^{#2}}
\newcommand{\ainv}[2]{\mathcal{A}_{#1}^{#2}}
\newcommand{\trapconstraint}[1]{\Theta({#1})}
\newcommand{\deadconstraint}[1]{\Delta({#1})}

\newcommand{\ils}{$\mathsf{IL1S}$}

\newcommand{\wss}{$\mathsf{WS1S}$}
\newcommand{\wssomega}{$\mathsf{WS}\omega\mathsf{S}$}
\newcommand{\mso}{$\mathsf{MSO}$}
\newcommand{\sils}{\mathsf{s}^n_{\scriptscriptstyle\mathit{IL1S}}}
\newcommand{\swss}{\mathsf{s}_{\scriptscriptstyle\mathit{WS1S}}}
\newcommand{\astruct}{\mathcal{S}}
\newcommand{\zero}{\bar{\mathsf{0}}}
\newcommand{\dom}{\mathrm{dom}}
\newcommand{\strord}{\sqsubseteq}
\newcommand{\wordeq}{\preceq}

\renewcommand{\succ}{\mathrm{succ}}
\newcommand{\apred}{\mathsf{pr}}

\newcommand{\dualnopar}[1]{{#1}^{\sim}}
\newcommand{\dual}[1]{\left({#1}\right)^{\sim}}

\newcommand{\minsem}[1]{\sem{#1}^{\mathrm{min}}}
\newcommand{\nsem}[1]{\sem{#1}^n}
\newcommand{\minequiv}{\equiv^{\mathrm{min}}}

\newcommand{\bool}[2]{\mathrm{B}_{{#1}}\!\left({#2}\right)}

\renewcommand{\proof}[1]{\ifLongVersion \noindent\emph{Proof}: {#1} \vspace*{\baselineskip}\else\fi}

%% file: abstract.tex
We consider concurrent systems consisting of a finite but unknown
number of components, that are replicated instances of a given set of
finite state automata. The components communicate by executing
interactions which are simultaneous atomic state changes of a set of
components. We specify both the type of interactions (e.g.\
rendez-vous, broadcast) and the topology (i.e.\ architecture) of the
system (e.g.\ pipeline, ring) via a decidable interaction logic, which
is embedded in the classical weak sequential calculus of one successor
(\wss). Proving correctness of such system for safety properties, such
as deadlock freedom or mutual exclusion, requires the inference of an
inductive invariant that subsumes the set of reachable states and
avoids the unsafe states. Our method synthesizes such invariants
directly from the formula describing the interactions, without costly
fixed point iterations. We applied our technique to the verification
of several textbook examples, such as dining philosophers, mutual
exclusion protocols and concurrent systems with preemption and
priorities.

%% file: body.tex
\section{Introduction}

The problem of parametric verification asks whether a system composed
of $n$ replicated processes is safe, for all $n \geq 2$. By safety we
mean that every execution of the system stays clear of a set of global
error configurations, such as deadlocks or mutual exclusion
violations. Even if we assume each process to be finite-state and
every interaction to be a synchronization of actions without data
exchange, the problem remains challenging because we want a general
proof of safety, that works for any number of processes.

In general, parametric verification is undecidable if unbounded data
is exchanged \cite{AptKozen86}, while various restrictions of
communication (rendez-vous) and architecture\footnote{We use the term
  architecture for the shape of the graph along which the interactions
  take place.} (ring, clique) define decidable subproblems
\cite{ClarkeGrumbergBrowne86,GermanSistla92,EmersonNamjoshi95,AminofKotekRubinSpegniVeith17}.
\ifLongVersion
Seminal works consider \emph{rendez-vous} communication, allowing a
fixed number of participants
\cite{ClarkeGrumbergBrowne86,GermanSistla92,EmersonNamjoshi95}, placed
in a ring \cite{ClarkeGrumbergBrowne86,EmersonNamjoshi95} or a clique
\cite{GermanSistla92}. Recently, \mso-definable graphs (with bounded
tree- and cliquewidth) and point-to-point rendez-vous communication
were considered in \cite{AminofKotekRubinSpegniVeith17}.

Most approaches to decidability focus on computing a \emph{cut-off}
bound $c$, that reduces the verification problem from $n \geq 2$ to at
most $c$ processes
\cite{ClarkeGrumbergBrowne86,EmersonNamjoshi95}. Other methods
identify systems with well-structured transition relations, for which
symbolic enumeration of reachable states is feasible \cite{Abdulla10}
or reduce to known decidable problems, such as reachability in vector
addition systems \cite{GermanSistla92}.
\fi
When theoretical decidability is not of concern, semi-algorithmic
techniques such as \emph{regular model checking}
\cite{KestenMalerMarcusPnueliShahar01,AbdullaHendaDelzannoRezine07},
SMT-based \emph{bounded model checking}
\cite{AlbertiGhilardiSharygina14,ConchonGoelKrsticMebsoutZaidi12},
\emph{abstraction}
\cite{BaukusBensalemLakhnechStahl00,BouajjaniHabermehlVojnar04} and
\emph{automata learning} \cite{ChenHongLinRummer17} can be used to
deal with more general classes of systems. An exhaustive chart of
existing parametric verification techniques is drawn in
\cite{BloemJacobsKhalimovKonnovRubinVeithWidder15}.

The efficiency of a semi-algorithmic method crucially relies on its
ability of synthesizing an \emph{inductive safety invariant}, that is
an infinite set of global configurations, which contains the initial
configurations, is closed under the transition relation, and excludes
the error configurations. In general, automatically synthesizing
invariants requires computationally expensive fixpoint iterations
\cite{CousotCousot79}. In the particular case of parametric systems,
invariants can be either \emph{global}, relating the local states of
all processes \cite{DamsLakhnechSteffen01}, or \emph{modular},
relating the states of few processes, of unimportant identities
\cite{PnueliRuahZuck01,ClarkeTalupurVeith06}.

We focus on parametric systems described using the
Behavior-Inter\-action-Priorities (BIP) framework \cite{BasuBBCJNS11},
in which processes are instances of finite-state \emph{component
  types}, whose interfaces are sets of \emph{ports}, labeling
transitions between local states, and interactions are sets of
strongly synchronizing ports, described by formulae of an
\emph{interaction logic}. An interaction formula captures the
architecture of the interactions (pipeline, ring, clique, tree) and
the communication scheme (rendez-vous, broadcast), which are not
hardcoded, but rather specified by the system designer.

As a distinguishing feature, we synthesize invariants directly from
the interaction formula of a system, without iterating its transition
relation. Such invariants depend only on the structure (and not on the
operational semantics) of the interaction network, described by a
Petri Net of unbounded size, being thus \emph{structural}
invariants. Essentially, the invariants we infer use the
\emph{traps}\footnote{Called in this way by analogy with the notion of
  traps for Petri Nets \cite{Sifakis78}.} of the system, which are
sets $W$ of local states with the property that, if a process is in a
state from $W$ initially, then always some process will be in a state
from $W$. We call these invariants \emph{trap invariants}
\cite{BensalemBNS09,BozgaIosifSifakis19a}.

Infering trap invariants from interaction formulae relies on two
logical operations: \begin{inparaenum}[(a)]
\item the \emph{positivation} operation, producing a weaker formula
  with the same minimal models, and
\item the \emph{dualization} operation, that essentially switches the
  conjunctions with disjunctions and the universal with existential
  quantifiers.
\end{inparaenum}
Although dualization is just a linear time syntactic transformation of
formulae, positivation is a more involved operation, depending on the
semantics of the underlying logic. A definition of positivation for a
simple interaction logic, relying on equalities and disequalities
between process indices to describe clique architectures, is provided
in \cite{BozgaIosifSifakis19a}.

\paragraph{\em Our Contribution}
This paper describes a non-trivial a generalization of the method from
\cite{BozgaIosifSifakis19a}, that considers an interaction logic with
equality and uninterpreted monadic predicate symbols, which is
embedded into the combined theory of sets and Presburger cardinality
constraints \cite{KuncakNR06}. In addition, here we introduce a cyclic
(modulo-$n$, where $n$ is the unbounded parameter of the system)
successor function and embed our logic in the weak monadic logic of
one successor (\wss), for which validity of a formula boils down to
proving language emptiness of a finite Rabin-Scott automaton built
from that formula. This new logic naturally describes systems with
ring and pipeline, as well as previously considered clique/multiset
architectures. Moreover, we provide an example showing that the method
can be easily generalized to handle tree-like architectures.

The trap invariants method is incomplete, meaning that there exists
parametric systems that are safe for any number of components, but
whose trap invariant does not suffice to prove safety. We deal with
this problem by computing universal Ashcroft invariants
\cite{Ashcroft75}, able to add extra constraints inferred by
restricting the interaction formula of the parametric system to a
fixed set of symbolic components. This technique is orthogonal to the
trap invariant computation and resembles the computation of
\emph{invisible invariants} \cite{PnueliRuahZuck01}, but tailored to
the BIP framework we have chosen to work with.

\begin{wrapfigure}{r}{0.5\textwidth}
  \vspace*{-2\baselineskip}
  \begin{center}
    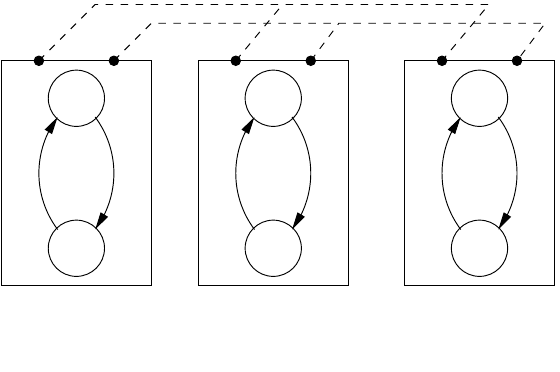
  \end{center}
  \vspace*{-\baselineskip}
  \caption{Parametric Dining Philosophers}
  \label{fig:philosophers}
  \vspace*{-2\baselineskip}
\end{wrapfigure}

\paragraph{\em Running Example}
Consider the dining philosophers system in
Fig. \ref{fig:philosophers}, consisting of $n\geq2$ components of type
$\mathsf{Fork}$ and $\mathsf{Philosopher}$ respectively, placed in a
ring of size $2n$. The $k$-th philosopher has a left fork, of index
$k$ and a right fork, of index $(k+1) \mod n$. Each component is an
instance of a finite state automaton with states $f$(ree) and $b$(usy)
for $\mathsf{Fork}$, respectively $w$(aiting) and $e$(ating) for
$\mathsf{Philosopher}$. A fork goes from state $f$ to $b$ via a
$t$(ake) transition and from $f$ to $b$ via a $\ell$(leave)
transition. A philosopher goes from $w$ to $b$ via a $g$(et)
transition and from $e$ to $w$ via a $p$(ut) transition. In this
example, we assume that the $g$ action of the $k$-th philosopher is
executed jointly with the $t$ actions of the $k$-th and $(k+1) \mod n$
forks, in other words, the philosopher takes both its left and right
forks simultaneously. Similarly, the $p$ action of the $k$-th
philosopher is executed simultaneously with the $\ell$ action of the
$k$-th and $[(k+1) \mod n]$-th forks, i.e.\ each philosopher leaves
both its left and right forks at the same time. We describe the
interactions of the system by the following first order formula
\(\interform_{\mathit{philo}} = \exists i ~.~ [g(i) \wedge t(i) \wedge
  t(s(i))] \vee [p(i) \wedge \ell(i) \wedge \ell(\succ(i))]\), where
transition labels (ports) are encoded as monadic predicate symbols and
$\succ(.)$ is the function symbol which denotes the successor of an
index in the ring. Each interaction is defined by a model of this
formula, for instance the structure interpreting $g$ as the set
$\set{k}$ and $t$ as the set $\set{k,(k+1)\mod n}$ corresponds to the
interaction of the $k$-th philosopher taking its forks, where $0 \leq
k < n$ is some index. The ring topology is implicit in the modulo-$n$
interpretation of the successor function $s$ as each $k$-th component
interacts with its $k$-th and $\succ(k)$-th neighbours only.


Intuitively, the system is deadlock-free for any $n\geq2$ since there
is no circular waiting scenario involving all the philosophers at
once. A rigorous proof requires an invariant disjoint from the set of
deadlock states, defined by the formula
\(\deadconstraint{\interform_{\mathit{philo}}} = \forall i ~.~ [\neg
  w(i) \vee \neg f(i) \vee \neg f(\succ(i))] \wedge [\neg e(i) \vee
  \neg b(i) \vee \neg b(\succ(i))]\). Our method computes a trap
invariant corresponding to the set of solutions of the following
constraint \(\trapconstraint{\interform_{\mathit{philo}}} = \forall i
~.~ w(i) \vee f(i) \vee f(\succ(i)) \leftrightarrow e(i) \vee b(i)
\vee b(\succ(i))\), derived from the interaction formula $\interform$
and the local structure of the component types.  Together with an
automata-based decision procedure for the interaction logic, this
invariant allows to prove deadlock freedom for the system in
Fig. \ref{fig:philosophers} in $\sim\!\!0.1$ seconds on an average
machine.

\section{Parametric Component-based Systems}

A \emph{component type} is a tuple $\acomptype = \tuple{\ports,
  \states, \initstate, \rules}$, where $\ports = \set{p,q,r,\ldots}$
is a finite set of \emph{ports}, $\states$ is a finite set of
\emph{states}, $\initstate\in\states$ is an initial state and $\rules
\subseteq \states \times \ports \times \states$ is a set of
\emph{transitions} $s \arrow{p}{} s'$. To simplify the upcoming
technical details, we assume there are no two different transitions
with the same port and write, for a port $p$ that labels a transition
$s \arrow{p}{} s' \in \rules$, $\pre{p}$ and $\post{p}$ for the source
$s$ and destination $s'$ of that (unique) transition, respectively.

A \emph{component-based system} $\asys =
\tuple{\typeno{\acomptype}{1}, \ldots,
  \typeno{\acomptype}{K},\interform}$ consists of a fixed number ($K$)
of component types $\typeno{\acomptype}{k} =
\tuple{\typeno{\ports}{k}, \typeno{\states}{k},
  \typeno{\initstate}{k}, \typeno{\rules}{k}}$ and an
\emph{interaction formula} $\interform$. 
\ifLongVersion
We shall sometimes write $\ports(\typeno{\acomptype}{k})$,
$\states(\typeno{\acomptype}{k})$,
$\initstate(\typeno{\acomptype}{k})$ and
$\rules(\typeno{\acomptype}{k})$ for $\typeno{\ports}{k}$,
$\typeno{\states}{k}$, $\typeno{\initstate}{k}$ and
$\typeno{\rules}{k}$, respectively.
\fi
Without loss of generality, we assume that $\typeno{\ports}{i} \cap
\typeno{\ports}{j} = \emptyset$, for all $1 \leq i < j \leq K$ and
unambiguously denote by $\typeof{p} \isdef \typeno{\acomptype}{k}$ the
component type of a port $p \in \typeno{\ports}{k}$. For instance, in
Fig. \ref{fig:philosophers} we have $\typeof{g\ell} = \typeof{gr} =
\typeof{p} = \mathsf{Philosopher}$ and $\typeof{g} = \typeof{\ell} =
\mathsf{Fork}$.

The interaction logic intentionally uses the names of the ports and
states, here viewed as monadic predicate symbols $\preds =
\bigcup_{k=1}^K (\typeno{\ports}{k} \cup \typeno{\states}{k})$, where
$\apred \in \preds$ is an arbitrary predicate symbol. In addition, we
consider a countable set $\vars$ of first-order variables and a set of
constant symbols $\const$. The only function symbol of arity greater
than zero is $\succ(.)$, denoting the \emph{successor}
function. Interaction formulae are written in the syntax of
\emph{Interaction Logic with One Successor} (\ils), given below:
\[\begin{array}{rclr}
t & := & x \in \vars \mid c \in \const \mid \succ(t) & \text{ terms} \\
\phi & := & t_1 \leq t_2 \mid \apred(t) \mid \phi_1 \wedge \phi_2 \mid \neg\phi_1 \mid \exists x ~.~ \phi_1 & \text{ formulae}
\end{array}\]
A \emph{sentence} is a formula in which all variables are in the scope
of a quantifier. A formula is \emph{positive} if each predicate symbol
occurs under an even number of negations and \emph{ground} if it
contains no variables. We write $t_1 = t_2$ for $t_1 \leq t_2 \wedge
t_2 \leq t_1$, $\phi_1 \vee \phi_2$ for $\neg(\neg\phi_1 \wedge
\neg\phi_2)$, $\phi_1 \rightarrow \phi_2$ for $\neg\phi_1 \vee
\phi_2$, $\phi_1 \leftrightarrow \phi_2$ for $\phi_1 \rightarrow
\phi_2 \wedge \phi_2 \rightarrow \phi_1$, $\forall x ~.~ \phi$ for
$\neg\exists ~.~ \neg\phi$.

For a positive integer $n>0$, we denote by $[n]$ the set
$\set{0,\ldots,n-1}$. We interpret \ils\ formulae over structures $\I
= ([n],\iota,\nu)$, where $[n]$ is the universe, $\iota : \const \cup
\preds \rightarrow [n] \cup 2^{[n]}$ maps constant symbols into
elements and predicate symbols into subsets of $[n]$, respectively,
and $\nu : \vars \rightarrow [n]$ maps variables into elements of
$[n]$. The successor function symbol $\succ(.)$ is always interpreted
in $\I$ by the function $\sils(x) = (x+1) \mod n$ and the inequality
relation by the set $\set{(u,v) \in [n] \times [n] \mid u \leq v}$.
With these definitions, the truth value of a formula $\phi$ in $\I$ is
defined recursively on the structure of $\phi$ and we write $\I
\models \phi$ when this value is true.

\paragraph{\em Remark~}
We adopted a circular modulo-$n$ interpretation of the successor
function, to naturally accomodate ring-like architectures, common in
distributed system design practice. This is not a restriction, because
clique architectures, where every component can interact with any
other, can be described using only equality and disequality
atoms. Moreover, acyclic pipeline architectures can be described using
the order relation, as follows: we identify a least and a greatest
element in the domain, namely \(\mathit{inf}(x) \isdef \forall y ~.~ x
\leq y\) and \(\mathit{sup}(x) \isdef \forall y ~.~ y \leq x\), and
describe interactions only for indices that are between those
elements. For the set of indices $x$ such that $\exists \zeta \exists
\xi ~.~ \mathit{inf}(\zeta) \wedge \mathit{sup}(\xi) \wedge \zeta \leq
x \wedge x < \xi$ holds, the graph of the successor function is
acyclic. Further, in \S\ref{sec:ils-wss} we show that \ils\ can be
extended with equalities modulo constants, such as the \emph{even} and
\emph{odd} predicates, without changing the invariant synthesis method
upon which our verification technique is based. Finally, in
\S\ref{sec:trap-automata}, we argue that tree architectures can be
fitted in this framework, with minimal changes. This claim is
sustained by an example in
\S\ref{sec:experiments}. \hfill$\blacksquare$

One of the consequences of the modulo-$n$ interpretation of the
successor function symbol is the existence of a \ils~ formula that
states the exact cardinality of the model: $\exists x ~.~ \succ^k(x)=x
\wedge \bigwedge_{i=1}^{k-1} \neg \succ^i(x)=x$. This formula is true
if and only if the cardinality of the universe equals the constant
$k$.  Since the purpose of \ils~ is to specify interactions in a
system whose number of components is arbitrary, we shall restrict
interaction formulae to finite disjunctions of formulae of the form
below:
\begin{equation}\label{eq:param-interform}
 \begin{array}{c}
   \exists x_1 \ldots \exists x_\ell ~.~ \varphi \wedge
   \bigwedge_{j=1}^\ell p_j(x_j) \wedge \bigwedge_{j=\ell+1}^{\ell+m}
   \forall x_j ~.~ \psi_j \rightarrow p_j(x_j)
 \end{array}
\end{equation}
where $\varphi,\psi_{\ell+1}, \ldots, \psi_{\ell+m}$ are conjunctions
of inequalities involving index variables, such that no comparison
between terms with the same variable is allowed, i.e.\ $\varphi$ and
$\psi_j$ do not contain atomic propositions of the form $\succ^i(x)
\leq \succ^j(x)$ for $i,j > 0$. Moreover, we assume that $\typeof{p_i}
= \typeof{p_j} \Rightarrow p_i = p_j$, for all $1 \leq i < j \leq
\ell+m$, i.e.\ the formula does not specify interactions between
different ports of the same component type\footnote{This restriction
  simplifies the technical presentation of the results and can be
  removed w.l.o.g.}.

Informally, the formula (\ref{eq:param-interform}) states that at most
$\ell$ components can simultaneously engage in a multiparty
rendez-vous, together with a broadcast to the ports $p_{\ell+1},
\ldots, p_{\ell+m}$ of the components whose indices satisfy the
constraints $\psi_{\ell+1}, \ldots, \psi_{\ell+m}$, respectively. An
example of peer-to-peer rendez-vous with no broadcast is the dining
philosophers system in Fig. \ref{fig:philosophers}, whereas examples
of broadcast are found among the test cases in
\S\ref{sec:experiments}.


\subsection{Execution Semantics of Component-based Systems}

The semantics of a component-based system is defined by a 1-safe Petri
Net, whose (reachable) markings and actions characterize the
(reachable) global states and transitions of the system. For reasons
of self-completeness, we recall below several basic definitions
relative to Petri Nets.

Formally, a \emph{Petri Net} (PN) is a tuple $\anet =
\tuple{\places,\trans,\edges}$, where $\places$ is a set of
\emph{places}, $\trans$ is a set of \emph{transitions}, $\places \cap
\trans = \emptyset$, and $\edges \subseteq \places \times \trans \cup
\trans \times \places$ is a set of \emph{edges}. The elements of
$\places \cup \trans$ are called \emph{nodes}. Given nodes $x,y \in
\places \cup \trans$, we write $E(x,y)\isdef1$ if $(x,y) \in E$ and
$E(x,y)\isdef0$, otherwise. For a node $x$, let $\pre{x} \isdef \set{y
  \in \places \cup \trans \mid E(y,x)=1}$, $\post{x} \isdef \set{y \in
  \places \cup \trans \mid E(x,y)=1}$ and lift these definitions to
sets of nodes, as usual.

A \emph{marking} of $\anet$ is a function $\amark : \places
\rightarrow \nat$. A transition $t$ is \emph{enabled} in $\amark$ if
and only if $\amark(s) > 0$ for each place $s \in \pre{t}$. The
\emph{transition relation} of $\anet$ is defined as follows. For all
markings $\amark$, $\amark'$ and all transitions $t$, we write $\amark
\arrow{t}{} \amark'$ whenever $t$ is enabled in $\amark$ and
$\amark'(s) = \amark(s) - E(s,t) + E(t,s)$, for all $s \in
\places$. Given two markings $\amark$ and $\amark'$, a finite sequence
of transitions $\sigma = t_1, \ldots,t_n$ is a \emph{firing sequence},
written $\amark \arrow{\sigma}{} \amark'$ if and only if
either \begin{inparaenum}[(i)]
\item $n=0$ and $\amark=\amark'$, or
\item $n\geq1$ and there exist markings $\amark_1, \ldots,
  \amark_{n-1}$ such that $\amark \arrow{t_1}{} \amark_1 \ldots
  \amark_{n-1} \arrow{t_{n}}{} \amark'$.
\end{inparaenum}

A \emph{marked Petri net} is a pair $\amarkednet=(\anet,\amark_0)$,
where $\amark_0$ is the \emph{initial marking} of $\anet$. A marking
$\amark$ is \emph{reachable} in $\amarkednet$ if and only if there
exists a firing sequence $\sigma$ such that $\amark_0 \arrow{\sigma}{}
\amark$. We denote by $\reach{\amarkednet}$ the set of reachable
markings of $\amarkednet$. A set of markings $\mathcal{M}$ is an
\emph{invariant} of $\amarkednet=(\anet,\amark_0)$ if and only if
$\amark_0 \in \mathcal{M}$ and for each $\amark \arrow{t}{} \amark'$
such that $\amark \in \mathcal{M}$, we have $\amark' \in
\mathcal{M}$. A marked PN $\amarkednet$ is \emph{$1$-safe} if
$\amark(s) \leq 1$, for each $s \in \places$ and $\amark \in
\reach{\amarkednet}$. All PNs considered in the following will be
1-safe and we shall silently blur the distinction between a marking
$\amark : \places \rightarrow \set{0,1}$ and the valuation $\nu_\amark
: \places \rightarrow \set{\bot,\top}$ defined as $\nu_\amark(s)=\top
\iff \amark(s)=1$.

Turning back to the definition of the semantics of component-based
parametric systems, let $\asys = \tuple{\typeno{\acomptype}{1},
  \ldots, \typeno{\acomptype}{K},\interform}$ be a system with
component types $\typeno{\acomptype}{k} = \tuple{\typeno{\ports}{k},
  \typeno{\states}{k}, \typeno{\initstate}{k}, \typeno{\rules}{k}}$,
for all $k=1,\ldots,K$. For each parameter $n\geq1$, we define a
marked PN $\amarkednet^n_\asys$, of size $\bigO(n)$, that
characterizes the set of executions of the instance of $\asys$ having
$n$ replicas of each component type. Formally, given a positive
integer $n\geq1$, we have $\amarkednet^n_\asys = (\anet,\amark_0)$,
where $\anet \isdef \tuple{\bigcup_{k=1}^K \typeno{\states}{k} \times
  [n],\trans,\edges}$ and whose sets of transitions $T$ and edges $E$
are defined from the interaction formula $\interform$, as follows.

\begin{figure}[htb]
  \vspace*{-\baselineskip}
  \begin{center}
    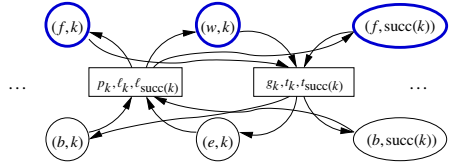
  \end{center}
  \vspace*{-\baselineskip}
  \caption{Unbounded Marked Petri Net for the Dining Philosophers Example}
  \label{fig:philosophers-pn}
  \vspace*{-\baselineskip}
\end{figure}

First, we define the set of minimal models of $\interform$, where
minimality is with respect to the pointwise inclusion of the sets that
interpret the predicate symbols. Formally, given structures
$\astruct_1 = ([n],\nu_1,\iota_1)$ and $\astruct_2 =
([n],\nu_2,\iota_2)$ sharing the same universe $[n]$, we have
$\astruct_1 \strord \astruct_2$ if and only if $\iota_1(\apred)
\subseteq \iota_2(\apred)$, for all $\apred \in \preds$. Given a
formula $\phi$, a structure $\astruct$ is a \emph{minimal model} of
$\phi$ if $\astruct \models \phi$ and, for all structures $\astruct'$
such that $\astruct' \strord \astruct$ and $\astruct'\neq\astruct$, we
have $\astruct' \not\models \phi$. We denote by $\minsem{\phi}$ the
set of minimal models of $\phi$. Two formulae $\phi_1$ and $\phi_2$
are \emph{minimally equivalent}, written as $\phi_1 \minequiv \phi_2$,
if and only if $\minsem{\phi_1} = \minsem{\phi_2}$.

Back to the definition of $\amarkednet^n_\asys$, for each minimal
model $\I = ([n],\nu,\iota) \in \minsem{\interform}$, we have a
transition $\atrans_\I \in T$ and edges $((s,i),\atrans_\I),
(\atrans_\I,(s',i)) \in E$, for all $s \arrow{p}{} s' \in
\bigcup_{k=1}^K \typeno{\rules}{k}$, such that $i \in \iota(p)$, and
nothing else is in $T$ or $E$. The initial marking of
$\amarkednet_\asys^n$ corresponds to the initial state of each
components, formally for each $1 \leq k \leq K$, each $s \in
\typeno{\states}{k}$ and each $1 \leq i \leq n$, $\amark_0((s,i)) = 1$
if $s = \typeno{\initstate}{k}$ and $\amark_0((s,i)) = 0$, otherwise.
For instance, Fig. \ref{fig:philosophers-pn} shows the PN for the
system in Fig. \ref{fig:philosophers}, with the initial marking
highlighted.

\ifLongVersion

Below we give a property of the marked PNs that define the semantics
of parametric component-based systems. 

\begin{definition}\label{def:decomposable-pn}
Given a component-based system $\asys$, a marked PN
$\amarkednet=(\anet, \amark_0)$, with $\anet=(\places, \trans,
\edges)$, is \emph{$\asys$-decomposable} if and only if there exists
an integer $n>0$ such that $\places = \bigcup_{k=1}^K
\typeno{\states}{k} \times [n]$ and in every reachable marking $\amark
\in \reach{\amarkednet}$, for each $1 \leq i \leq n$ and each $1 \leq
k \leq K$ there exists exactly one state $s \in \typeno{\states}{k}$
such that $\amark((s,i))=1$.
\end{definition}

\begin{lemma}\label{lemma:decomposable-pn}
  The marked PN $\amarkednet_\asys^n$ is $\asys$-decomposable, for
  each component-based system $\asys$ and each integer $n > 0$.
\end{lemma}
\proof{ Let $\asys = \tuple{\typeno{\acomptype}{1}, \ldots,
    \typeno{\acomptype}{K},\interform}$ be a system with component
  types $\typeno{\acomptype}{k} = \tuple{\typeno{\ports}{k},
    \typeno{\states}{k}, \typeno{\initstate}{k}, \typeno{\rules}{k}}$,
  for all $k=1,\ldots,K$, and let $n > 0$ be a parameter. Let
  $\amarkednet_\asys^n = (\anet, \amark_0)$ and $\amark \in
  \reach{\amarkednet_\asys^n}$ be a reachable marking. Then $\anet =
  (\bigcup_{k=1}^K \typeno{\states}{k} \times [n], \trans,
  \edges)$. We prove the property by induction on the length $\ell$ of
  the shortest path from $\amark_0$ to $\amark$. If $\ell=0$ the
  property holds because each component type $1 \leq k \leq K$ has
  exactly one initial state $\typeno{\initstate}{k}$ and only the
  states $(\typeno{\initstate}{k}, i)$ are initially marked, for all
  $1 \leq i \leq n$. For the induction step $\ell>0$, assume that
  $\amark' \arrow{\atrans}{} \amark$ and the property of Definition
  \ref{def:decomposable-pn} holds for $\amark'$. Then there exists
  $\I = ([n],\nu,\iota,\mu) \in \minsem{\interform}$, such that
  $\atrans = \atrans_\I$ and, for each $i \in [n]$ and each $p \in
  \typeno{\ports}{k}$ such that $s' \arrow{p}{} s \in
  \typeno{\rules}{k}$ and $i \in \iota(p)$, there are edges
  $((s',i),\atrans_\I), (\atrans_\I,(s,i)) \in \edges$. Suppose, for a
  contradiction, that there exists $1 \leq i_0 \leq n$ and $1 \leq k_0
  \leq K$ such that $\amark((s,i_0)) = \amark((s'',i_0)) = 1$, for two
  distinct states $s,s'' \in \typeno{\states}{k}$. Then
  $(\atrans_\I,(s,i_0)), (\atrans_\I,(s'',i_0)) \in \edges$ and $i_0
  \in \iota(p) \cap \iota(q)$, for two transition rules $s'
  \arrow{p}{} s, s' \arrow{q}{} s'' \in \typeno{\rules}{k}$. However,
  this comes in contradiction with the assumption that a transition
  does not involve two different ports from the same component type
  (\ref{eq:param-interform}). \qed}

\fi

\section{Computing Trap Invariants}
\label{sec:trap-invariants}

We leverage from a standard notion in the theory of Petri Nets to
define a class of invariants, that are useful for proving certain
safety properties. Given a Petri Net $\anet = (\places, \trans,
\edges)$, a set of places $W \subseteq \places$ is called a
\emph{trap} if and only if $\post{W} \subseteq \pre{W}$. A trap $W$ of
$\anet$ is an \emph{initially marked trap} (IMT) of the marked PN
$\amarkednet = (\anet,\amark_0)$ if and only if $\amark_0(s)=\true$
for some $s \in W$. An IMT of $\amarkednet$ is minimal if none of its
nonempty strict subsets is an IMT of $\amarkednet$. We denote by
$\imt{\amarkednet} \subseteq 2^\places$ the set of IMTs of
$\amarkednet$.

\begin{example}\label{ex:traps}
  \vspace*{-0.5\baselineskip}
  Consider an instance of the marked PN in
  Fig. \ref{fig:philosophers-pn} for $n=2$. For simplicity, we denote
  places $(f,k)$, $(w,k)$, $(b,k)$ and $(e,k)$ as $f_k, w_k, b_k$ and
  $e_k$ for $k=0,1$, respectively. The local states of each component
  form a minimal trap, i.e.\ $\set{f_k, b_k}$ and $\set{w_k,e_k}$ are
  traps, for $k=0,1$. In addition, $\set{w_0,e_0,w_1},
  \set{w_0,b_1,w_1}, \set{f_0,b_0,e_1}$ and $\set{f_1,b_0,e_1}$ are
  also minimal traps. \hfill$\blacksquare$
  \vspace*{-0.5\baselineskip}
\end{example}

An IMT defines an invariant of the PN, because some place in the trap
will always be marked, no matter which transition is fired. The
\emph{trap invariant} of $\amarkednet$ is the set of markings that
mark each IMT of $\amarkednet$. The trap invariant of $\amarkednet$
subsumes the set of reachable markings of $\amarkednet$, because the
latter is the least invariant of $\amarkednet$\footnote{Since
  invariants are closed under intersection, the least invariant is
  unique.}. To prove that a certain set of markings is unreachable, it
is sufficient to prove that the this set has empty intersection with
the trap invariant. For self-completeness, we briefly discuss the
computation of trap invariants for a given marked PN of fixed size,
before explaining how this can be done for marked PNs defining the
executions of parametric systems, which are of unknown sizes.

\begin{definition}\label{def:trap-constraint}
The \emph{trap constraint} of a PN $\anet = (\places,\trans,\edges)$
is the formula:
\[\begin{array}{c}
\trapconstraint{\anet} \isdef \bigwedge_{t \in T} \big(\bigvee_{\text{$x \in \pre{t}$}} x\big)
\rightarrow \big(\bigvee_{\text{$y \in \post{t}$}} y\big)
\end{array}\]
where each place $x,y \in \places$ is viewed as a propositional variable. 
\end{definition}
It is not hard to show\footnote{See e.g.\ \cite{BarkaouiLemaire89} for
  a proof.} that any boolean valuation $\beta : \places \rightarrow
\set{\bot,\top}$ that satisfies the trap constraint
$\trapconstraint{\anet}$ defines a trap $W_\beta$ of $\anet$ in the
obvious sense $W_\beta = \set{s \in \places \mid \beta(s) =
  \top}$. Further, if $\amark_0 : \places \rightarrow \set{0,1}$ is
the initial marking of a 1-safe PN $\anet$ and $\mu_0 \isdef
\bigvee_{\amark_0(s)=1} s$ is a propositional formula, then each
minimal satisfying valuation of $\mu_0 \wedge \trapconstraint{\anet}$
defines a minimal IMT of $(\anet,\amark_0)$, where minimality of
boolean valuations is considered with respect to the usual partial
order $\beta_1 \preceq \beta_2 \iff \forall s \in \places ~.~
\beta_1(s) \rightarrow \beta_2(s)$.

Usually, computing invariants requires building a sequence of
underapproximants whose limit is the least fixed point of an
abstraction of the transition relation of the system
\cite{CousotCousot79}. This is however not the case with trap
invariants, that can be directly computed by looking at the structure
of the system, captured by the trap constraint, and to the initial
marking. To this end, we introduce two operations on propositional
formulae. First, given a propositional formula $\phi$, we denote by
$\pos{\phi}$ the result of deleting (i.e.\ replacing with $\top$) the
negative literals from the DNF of $\phi$. It is not hard to show that
$\phi \minequiv \pos{\phi}$, i.e.\ this transformation preserves the
minimal satisfying valuations of $\phi$. We call this operation
\emph{positivation}.

Second, let $\dualnopar{\phi}$ denote the result of replacing, in the
negation normal form of $\phi$, all conjunctions by disjunctions and
viceversa. Formally, assuming that $\phi$ is in NNF, let:
\[\dual{\phi_1 \wedge \phi_2} \isdef \dualnopar{\phi_1} \vee \dualnopar{\phi_2}
\hspace*{1cm}
\dual{\phi_1 \vee \phi_2} \isdef \dualnopar{\phi_1} \wedge \dualnopar{\phi_2}
\hspace*{1cm}
\dual{\neg s} \isdef \neg s
\hspace*{1cm} \dualnopar{s} \isdef s\] For any boolean valuation
$\beta$, we have $\beta \models \phi \iff \overline{\beta} \models
\neg(\dualnopar{\phi})$, where $\overline{\beta}(s) \isdef
\neg\beta(s)$ for each propositional variable $s$. This operation is
usually referred to as \emph{dualization}.

The following lemma gives a straightforward method to compute trap
invariants, logically defined by a CNF formula with positive literals
only, whose clauses correspond to the (enumeration of the elements of
the) traps. It is further showed that such a formula defines an
invariant of the finite marked PN:

\begin{lemma}\label{lemma:pn-trap-invariant}
  Given a marked PN $\amarkednet = (\anet,\amark_0)$, we have
  \(\alltrap{\amarkednet} \equiv \dual{\pos{(\mu_0 \wedge
      \trapconstraint{\anet})}}\), where \(\alltrap{\amarkednet}
  \isdef \bigwedge_{W \in \imt{\amarkednet}} \bigvee_{s \in W} s\) and
  $\mu_0 \isdef \bigvee_{\amark_0(s)=1} s$. Moreover,
  $\sem{\alltrap{\amarkednet}}$ is the trap invariant of
  $\amarkednet$.
\end{lemma}
\proof{ Let $\anet = (\places, \trans, \edges)$ and $W \subseteq
  \places$ be a trap of $\anet$. We have the following equivalences:
  \[\begin{array}{rcll}
  \post{W} & \subseteq & \pre{W} & \iff \\
  \bigwedge_{p \in \places} [p \in W & \rightarrow & \set{t \in \trans \mid (p,t) \in \edges} \subseteq \set{t \in \trans \mid \bigvee_{q \in \places} (t,q) \in \edges}] & \iff \\
  \bigwedge_{p \in \places} [p \in W & \rightarrow & (\bigwedge_{t \in \trans} p \in \pre{t} \rightarrow \bigvee_{q \in \places} q \in W \wedge q \in \post{t})] & \iff \\
  \bigwedge_{p \in \places} \bigwedge_{t \in \trans} (p \in W \wedge p \in \pre{t} & \rightarrow & \bigvee_{q \in \places} q \in W \wedge q \in \post{t}) & \iff \\
  \bigwedge_{t \in \trans} (\bigvee_{\text{$p \in \pre{t}$}} p \in W & \rightarrow & \bigvee_{q \in \post{t}} q \in W)
  \end{array}\]
  If we use propositional variables $p$ and $q$ to denote $p \in W$
  and $q \in W$, respectively, we obtain the trap constraint
  $\trapconstraint{\anet}$ from the last formula. Hence, any boolean
  valuation $\beta \in \sem{\mu_0 \wedge \trapconstraint{\anet}}$
  corresponds to an initially marked trap $W_\beta \isdef \set{p \in
    \places \mid \beta(p) = \top}$. Further, since $\mu_0 \wedge
  \trapconstraint{\anet}$ is a propositional formula, each satisfying
  valuation corresponds to a conjunctive clause of its DNF. Hence the
  set of propositional variables in each conjunctive clause in the DNF
  of $\pos{\mu_0 \wedge \trapconstraint{\anet}}$ corresponds to an IMT
  and, moreover, every IMT has a corresponding conjunctive
  clause. Thus $\alltrap{\amarkednet} \equiv \dual{\pos{\mu_0 \wedge
      \trapconstraint{\anet}}}$ follows. The second point follows
  directly from the definition $\alltrap{\amarkednet}$. \qed}

The computation of a trap invariant consists of the following
steps: \begin{inparaenum}[(1)]
\item convert the propositional formula $\mu_0 \wedge
  \trapconstraint{\anet}$ in DNF,
\item for each conjunctive clause, remove the negative literals and
\item dualize the result.
\end{inparaenum}
Importantly, the first two steps can be replaced by any transformation
on formulae whose result is a positive formula that is minimally
equivalent to the input, because only the minimal traps are important
for the trap invariant. Moreover, the negative literals do not occur
in the propositional definition of a set of places, which is why we
require the input of dualization to be a positive formula\footnote{If
  the DNF is $(p \wedge q) \vee (p \wedge \neg r)$, the dualization
  would give $(p \vee q) \wedge (p \vee \neg r)$. The first clause
  corresponds to the trap $\set{p,q}$ (either $p$ or $q$ is marked),
  but the second does not directly define a trap. However, by first
  removing the negative literals, we obtain the traps $\set{p,q}$ and
  $\set{r}$.}. These two properties of positivation constitute the
basis of the definition of positivation for quantified \ils\ formulae,
next in \S\ref{sec:trap-automata}.

In the rest of this section we focus on computing trap invariants for
1-safe marked PNs obtained from parametric systems consisting of
$\bigO(n)$ components, where $n\geq1$ is an unknown parameter. We
write parametric trap constraints using the same logic \ils, used to
describe interaction formulae. Namely, if $\interform$ is an
interaction formula consising of a disjunction of formulae of the form
(\ref{eq:param-interform}), then $\trapconstraint{\interform}$ is the
conjunction of formulae of the form below (\ref{eq:trap-constraint}),
one for each (\ref{eq:param-interform}) formula in the disjunction:
\begin{equation}\label{eq:trap-constraint}
  \begin{array}{r}
    \small{
      \forall x_1 \ldots \forall x_\ell ~.~ \varphi \wedge 
      \left[\bigvee_{j=1}^\ell \pre{p_j}(i_j) \vee \bigvee_{j=\ell+1}^{\ell+m}
        \exists x_j ~.~ \psi_j \wedge \pre{p_j}(i_j)\right] } \\
    \small{
      \rightarrow \bigvee_{j=1}^\ell \post{p_j}(i_j) \vee \bigvee_{j=\ell+1}^{\ell+m}
      \exists x_j ~.~ \psi_j \wedge \post{p_j}(i_j) }
  \end{array}
\end{equation}
where, for a port $p \in \typeno{\ports}{k}$ of some component type
$\typeno{\acomptype}{k}$, $\pre{p(x)}$ and $\post{p(x)}$ denote the
unique predicate atoms $s(x)$ and $s'(x)$, such that $s \arrow{p}{} s'
\in \typeno{\rules}{k}$ is the unique transition involving $p$, or
$\bot$ if there is no such rule. Note that
$\trapconstraint{\interform}$ is the generalization of the trap
constraint $\trapconstraint{\anet}$ for a given fixed size PN, to the
case of a parametric system described by an interaction formula
$\interform$. For instance, the trap constraint of the Dining
Philosophers example from Fig. \ref{fig:philosophers}, with
interaction formula $\interform_{\mathit{philo}} = \exists i ~.~ [g(i)
  \wedge t(i) \wedge t(\succ(i))] \vee [p(i) \wedge \ell(i) \wedge
  \ell(\succ(i))]$ is \(\trapconstraint{\interform_{\mathit{philo}}} =
\forall i ~.~ w(i) \vee f(i) \vee f(\succ(i)) \leftrightarrow e(i)
\vee b(i) \vee b(\succ(i))\).

In order to define a trap invariant computation method for parametric
systems described using \ils\ interaction formulae, we need
counterparts of the propositional positivation and dualization
operations, obtained as follows: \begin{inparaenum}[(1)]
\item we translate \ils\ trap constraints into equivalent formulae of
  weak monadic second order logic of one successor (\wss), and
\item we leverage from the standard automata theoretic two-way
  translation between \wss\ and finite Rabin-Scott automata to define
  positivation and dualization directly on automata.
\end{inparaenum}
For presentation purposes, we define first dualization on
\wss\ formulae, however for efficiency, our implementation applies it
on automata directly. We have not been able to define a semantic
equivalent of positivation as an operation on \wss\ formulae, thus we
need to work with automata for this purpose.

\subsection{From \ils\ to \wss}
\label{sec:ils-wss}

We introduce the standard second order logic \wss\ interpreted over
finite words, by considering an infinite countable set $\Vars$ of set
variables, denoted as $X,Y,\ldots$ in the following. The syntax of
\wss\  is the following:
\[\begin{array}{rclr}
t & := & \zero \mid x \mid \succ(t) & \text{ terms} \\
\phi & := & t_1 = t_2 \mid \apred(t) \mid X(t) \mid
\phi_1 \wedge \phi_2 \mid \neg\phi_1 \mid \exists x ~.~ \phi_1 \mid
\exists X ~.~ \phi_1 & \text{ formulae} 
\end{array}\]
Note that the syntax of \wss\ is the syntax of \ils, extended with the
constant symbol $\zero$, atoms $X(t)$ and monadic second order
quantifiers $\exists X ~.~ \phi$. As discussed below, we consider
w.l.o.g. equality atoms $t_1=t_2$ instead of inequalities $t_1\leq
t_2$ in \ils.

\wss\ formulae are interpreted over structures $\astruct = ([n],
\iota, \nu, \mu)$, where $\iota$ and $\nu$ are as for \ils~ and $\mu :
\Vars \rightarrow 2^{[n]}$ is an interpretation of the set
variables. Moreover, the constant symbol $\zero$ is interpreted as the
integer zero and the successor function is interpreted differently, by
the function \(\swss(x) \isdef x+1\) if \(x < n-1\) and \(\swss(n-1)
\isdef n-1\)\footnote{By classical convention, the successor on a
  finite domain is a total function that loops on the greatest element
  \cite[Example 2.10.3]{KhoussainovNerode}. }. Inequalities $t_1 \leq
t_2$ can be defined in the usual way, using second-order transitive
closure of the successor relation and $t_1 < t_2$ stands for $t_1 \leq
t_2 \wedge \neg t_1 = t_2$. Moreover, $\zero$ can be defined using
inequality and is considered as part of the syntax mainly for the
conciseness of the presentation.

Next, we define an embedding of \ils\ formulae into \wss. W.l.o.g. we
consider \ils\ formulae that have been previously flattened, i.e\ the
successor function occurs only within atomic propositions of the form
$x = \succ(y)$. Roughly, this is done by replacing each atomic
proposition $\succ^i(x)=y$ by the formula $\forall x_1 \ldots \forall
x_{i-1} ~.~ x_1=\succ(x) \wedge \bigwedge_{j=1}^{i-2}
x_{j+1}=\succ(x_j) \rightarrow \succ(x_{i-1})=y$, the result being a
formula $\phi_{\mathit{flat}}$ in which only atoms of the form
$s(x)=y$ occur. Moreover, any constant symbol $c \in \const$ from the
input \ils\ formula is replaced by a fresh free variable $x_c$. Let
$\mathit{Tr}(\phi) \isdef \exists \xi ~.~ \forall y ~.~ y \leq \xi
\wedge \mathit{tr}(\phi)$, where $\mathit{tr}(\phi)$ is defined
recursively on the structure of $\phi$:
\[\begin{array}{rclcrcl}
\mathit{tr}(\succ(x)=y) & \isdef & (x < \xi \wedge \succ(x)=y) \vee (x = \xi \wedge y = \zero) & \hspace*{2mm} &
\mathit{tr}(x \leq y) & \isdef & x \leq y \\
\mathit{tr}(\apred(x)) & \isdef & \apred(x) &&
\mathit{tr}(\phi_1 \wedge \phi_2) & \isdef & \mathit{tr}(\phi_1) \wedge \mathit{tr}(\phi_2) \\
\mathit{tr}(\neg\phi_1) & \isdef & \neg\mathit{tr}(\phi_1) && 
\mathit{tr}(\exists x ~.~ \phi_1) & \isdef & \exists x ~.~ \mathit{tr}(\phi_1)
\end{array}\] and $\xi$ is not among the free variables
of $\phi$.

\begin{lemma}\label{lemma:ils-wss}
  Given an \ils~ formula $\phi$, the following are
  equivalent: \begin{compactenum}
  \item\label{it1:ils-wss} $([n],\nu,\iota) \models \phi$, 
  \item\label{it2:ils-wss} $([n],\nu,\iota,\mu) \models
    \mathit{Tr}(\phi)$, for any $\mu : \Vars \rightarrow 2^{[n]}$.
  \end{compactenum}
\end{lemma}
\proof{ ``(\ref{it1:ils-wss}) $\Rightarrow$ (\ref{it2:ils-wss})''
  First, it is routine to check that, for any \wss-structure, we have
  $([n],\iota,\nu,\mu) \models \forall y ~.~ y \leq x \iff
  \nu(x)=n-1$. Suppose that $\phi$ has a model $\I=([n],\iota,\nu)$
  and the interpretation of $s$ is $\sils$. Then we show that
  $\astruct=([n],\iota,\nu,\mu)$ is a model of $\mathit{Tr}(\phi)$,
  for any $\mu : \Vars \rightarrow 2^{[n]}$. For this, it is enough to
  show that $([n],\iota,\nu[\mu\leftarrow n-1],\mu) \models
  \mathit{tr}(\phi)$, by induction on the structure of $\phi$. The
  base cases are: \begin{compactitem}
  \item $s(x)=y$: in this case $\sils(\nu(x))=\nu(y)$ and thus
    $\nu(y)=(\nu(x)+1) \mod n$, by the definition of $\sils$. But then
    either $\nu(x) < n-1$ and $\nu(y)=\nu(x)+1$ or $\nu(x)=n-1$ and
    $\nu(y)=0$, thus $\astruct \models \mathit{tr}(s(x)=y)$, as
    required.
  \item $\apred(x)$: in this case $\nu(x) \in \iota(p)$ and $\astruct
    \models \mathit{tr}(\apred(x))$ by the definition. 
  \end{compactitem}
  The induction cases are immediate.

  ``(\ref{it1:ils-wss}) $\Leftarrow$ (\ref{it2:ils-wss})'' If
  $([n],\nu,\iota,\mu) \models \mathit{Tr}(\phi)$, for some arbitrary
  mapping $\mu : \Vars \rightarrow 2^{[n]}$, then we have
  $([n],\nu[\nu \leftarrow n-1],\iota,\mu) \models \mathit{tr}(\phi)$
  and we show $([n],\nu,\iota) \models \phi$ by induction on the
  structure of $\phi$. The most interesting case is when $\phi$ is
  $s(x)=y$, in which case either: \begin{compactitem}
  \item $\nu(x) < n-1$ and $\nu(y)=\nu(x)+1$, or
  \item $\nu(x)=n-1$ and $\nu(y)=0$. 
  \end{compactitem}
  In each case, we have $\sils(\nu(x))=\nu(y)$, hence $([n],\nu,\iota)
  \models s(x)=y$, as required. \qed}

\paragraph{\em Remark~}
The above translation can be easily generalized to the case where
\ils\ contains any \wss-definable relation, such as the
$\mathit{even}(x)$ predicate, defined below:
\[\begin{array}{rcl}
\mathit{even}(x) & \isdef & \exists X \exists Y ~.~ X(x)
\wedge X(\zero) \wedge \forall y ~.~ X(y) \leftrightarrow \neg Y(y) \wedge \\
&& \forall y ~.~ X(y) \wedge y \neq \succ(y) \rightarrow Y(\succ(y)) \wedge
\forall y ~.~ Y(y) \wedge y \neq \succ(y) \rightarrow X(\succ(y))
\end{array}\]
Analogously, we can include any modulo constraint of the form $x
\equiv_k \ell$, where $k>0$ and $0 \leq \ell < k$ are integer
constants. \hfill$\blacksquare$

Next, we define the \emph{dualization} $\dualnopar{\phi}$ of a \wss\ 
formula $\phi$, in negative normal form:
\[\begin{array}{cc}
\begin{array}{rclcrcl}
\dual{t_1 = t_2} & \isdef & \neg t_1 = t_2 & ~ & \dual{\neg t_1 = t_2} & \isdef & t_1 = t_2 \\
\dualnopar{\apred(t)} & \isdef & \apred(t) && \dual{\neg\apred(t)} & \isdef & \neg\apred(t) \\
\dualnopar{X(t)} & \isdef & \neg X(t) && \dual{\neg X(t)} & \isdef & X(t) \\
\end{array}
&
\begin{array}{rclcrcl}
\dual{\phi_1 \wedge \phi_2} & \isdef & \dualnopar{\phi_1} \vee \dualnopar{\phi_2} & ~ & \dual{\phi_1 \vee \phi_2} & \isdef & \dualnopar{\phi_1} \wedge \dualnopar{\phi_2} \\
\dual{\exists x ~.~ \phi_1} & \isdef & \forall x ~.~ \dualnopar{\phi_1} && \dual{\forall x ~.~ \phi_1} & \isdef & \exists x ~.~ \dualnopar{\phi_1} \\
\dual{\exists X ~.~ \phi_1} & \isdef & \forall X ~.~ \dualnopar{\phi_1} && \dual{\forall X ~.~ \phi_1} & \isdef & \exists X ~.~ \dualnopar{\phi_1} 
\end{array}
\end{array}\]
Note that dualization acts differently on predicate literals of the
form $\apred(t)$ and $\neg\apred(t)$ than on literals involving a set
variable $X(t)$ and $\neg X(t)$. Namely, the former are left
unchanged, whereas the latter are negated. Its formal property is
stated below:

\begin{lemma}\label{lemma:wss-dual}
  Given a \wss\  formula $\phi$, for every structure
  $\astruct=([n],\nu,\iota,\mu)$ we have $\astruct \models \phi \iff
  \overline{\astruct} \models \neg(\dualnopar{\phi})$, where
  $\overline{\astruct} \isdef ([n],\nu,\overline{\iota},\mu)$ and for
  each $\apred \in \preds$, $\overline{\iota}(\apred) \isdef [n]
  \setminus \iota(\apred)$.
\end{lemma}
\proof{By induction on the structure of $\phi$: \begin{compactitem}
  \item $t_1=t_2$ and $\neg t_1=t_2$: the truth value of this atom is
    the same in $\astruct$ and $\overline{\astruct}$ and moreover
    $t_1=t_2$ and $\neg\dual{t_1=t_2}$ are equivalent.
  \item $X(t)$ and $\neg X(t)$: same as above.
  \item $\apred(t)$: the interpretation of $t$ is the same in
    $\astruct$ and $\overline{\astruct}$, because it depends only on
    $\nu$. Let $k \in [n]$ be this value. Then we obtain: \[\astruct \models
    \apred(t) \iff k \in \iota(\apred) \iff k \not\in
    \overline{\iota}(\apred) \iff \overline{\astruct} \models
    \neg\apred(t)\enspace.\]
  \item $\neg\apred(t)$: a consequence of the equivalence $\astruct
    \models \apred(t) \iff \overline{S} \models \neg\apred(t)$,
    established at the previous point.
  \end{compactitem}
  The rest of the cases are easy applications of the induction
  hypothesis.  \qed}

\ifLongVersion

For technical reasons, we also introduce a \emph{booleanization}
operation that, given a \wss\ formula $\phi$ and a positive constant
$n>0$, produces a propositional formula $\bool{n}{\phi}$ with the
property that each model $([n],\nu,\iota,\mu)$ of $\phi$ can be turned
into a satisfying boolean valuation for $\bool{n}{\phi}$ and
viceversa, from every boolean model of $\bool{n}{\phi}$ one can
extract a model of $\phi$.

First, given an integer $i \geq 0$ and a \wss\  formula $\phi(x)$, we
denote by $\phi[i/x]$ (resp. $t[i/x]$) the formula (term) obtained
from $\phi$ (resp. $t$) by replacing every occurrence of $x$ with the
term $s^i(\zero)$, where $s^i$ denotes $i$ successive applications of
the successor function. Second, for a set $S$ of positive integers,
the formula $\phi[S/X]$ is defined homomorphically, starting with the
base case $X(t)[S/X] \isdef \bigvee_{i \in S} t=s^i(\zero)$.

\[\begin{array}{rclcrcl}
\bool{n}{s^i(\zero) = s^j(\zero)} & \isdef & i=j \vee (i\geq n-1
\wedge j \geq n-1) & \hspace*{2mm} & \bool{n}{\apred(s^i(\zero))} &
\isdef & \apred_{\min(i,n-1)} \\ \bool{n}{\phi_1 \wedge \phi_2} &
\isdef & \bool{n}{\phi_1} \wedge \bool{n}{\phi_2} &&
\bool{n}{\neg\phi_1} & \isdef & \neg\bool{n}{\phi_1}
\\ \bool{n}{\exists x ~.~ \phi} & \isdef & \bigvee_{i \in [n]}
\bool{n}{\phi[i/x]} && \bool{n}{\exists X ~.~ \phi} & \isdef &
\bigvee_{S \subseteq [n]} \bool{n}{\phi[S/X]}
\end{array}\]
where, for any $\apred \in \preds$ and $j \in [n]$, $\apred_j$ is a
propositional variable ranging over the boolean values $\top$ (true)
and $\bot$ (false). Moreover, we relate \wss\  structures with boolean
valuations as follows. Given a structure
$\astruct=([n],\nu,\iota,\mu)$ we define the boolean valuation
$\beta_\astruct(\apred_j) \isdef \top \iff \swss^j(0) \in
\iota(\apred)$, for all $\apred \in \preds$ and $j \in [n]$. The
following lemma states the formal property of booleanization:

\begin{lemma}\label{lemma:wss-bool}
  Given a \wss\  sentence $\phi$ and $n>0$, for every structure
  $\astruct = ([n],\nu,\iota,\mu)$, we have $\astruct \models \phi
  \iff \beta_\astruct \models \bool{n}{\phi}$.
\end{lemma}
\proof{  
  We prove the following more general statement. Let
  $\phi(x_1,\ldots,x_k,X_1,\ldots,X_m)$ be a \wss\  formula with free
  variables $x_1,\ldots,x_k \in \vars$ and $X_1,\ldots,X_m \in \Vars$,
  $i_1,\ldots,i_k \in [n]$ and $S_1, \ldots, S_m \subseteq [n]$. Then
  we show that: \[\begin{array}{c}
  \astruct \models \phi[i_1/x_1, \ldots, i_k/x_k,S_1/X_1,\ldots,S_m/X_m] \\
  \iff \\
  \beta_\astruct \models \bool{n}{\phi[i_1/x_1, \ldots, i_k/x_k,S_1/X_1,\ldots,S_m/X_m]}
  \end{array}\] by
  induction on the structure of $\phi$: \begin{compactitem}
  \item $t_1 = t_2$: since $\phi[i_1/x_1, \ldots,
    i_k/x_k,S_1/X_1,\ldots,S_m/X_m]$ is a sentence, it must be the
    case that $t_1 = s^{i_1}(\zero)$ and $t_2 = s^{i_2}(\zero)$, for
    some $i_1,i_2 \geq 0$. Then we have:
    \[\begin{array}{rcl}
    \astruct \models s^{i_1}(\zero) = s^{i_2}(\zero) & \iff & \swss^{i_1}(0)=\swss^{i_2}(0) \\[2mm]
    & \iff & i_1 = i_2 \vee (i_1 \geq n-1 \wedge i_2 \geq n-1) \\[2mm]
    & \iff & \beta_\astruct \models \bool{n}{s^{i_1}(\zero) = s^{i_2}(\zero)}\enspace.
    \end{array}\]
  \item $\apred(t)$: since $\phi[i_1/x_1, \ldots,
    i_k/x_k,S_1/X_1,\ldots,S_m/X_m]$ is a sentence, it must be the
    case that $t = s^i(\zero)$, for some $i \geq 0$. We obtain:
    \[\begin{array}{rcl}
    \astruct \models \apred(s^i(\zero)) & \iff & \swss^{i}(\zero) \in \iota(\apred) \\[2mm]
    & \iff & \swss^{\min(i,n-1)} \in \iota(\apred) \\[2mm]
    & \iff & \beta_\astruct \models \apred_{\min(i,n-1)}\enspace.
    \end{array}\]
  \end{compactitem}
  The rest of the cases are easy applications of the induction hypothesis. 
  \qed}

Finally, we relate \wss\ dualization, booleanization and propositional
dualization:

\begin{lemma}\label{lemma:bool-dual}
  Given a \wss\  formula $\phi$ and an integer $n > 0$, we have
  $\bool{n}{\dualnopar{\phi}} \equiv \dualnopar{\bool{n}{\phi}}$.
\end{lemma}
\proof{ Let $\beta : \set{\apred_k \mid \apred \in \preds, k \in [n]}
  \rightarrow \set{\top,\bot}$ be an arbitrary boolean valuation and
  let $\astruct = ([n],\nu,\iota,\mu)$ be a structure such that, for
  each $\apred \in \preds$, we have $\iota(\apred) = \set{k \in [n]
    \mid \beta(\apred_k) = \top}$ and $\nu$, $\mu$ are picked at
  random. Obviously, we have that $\beta = \beta_\astruct$, hence by
  Lemma \ref{lemma:wss-bool}, $\beta \models
  \bool{n}{\dualnopar{\phi}} \iff \astruct \models \dualnopar{\phi}$
  and by Lemma \ref{lemma:wss-dual} we get $\astruct \models
  \dualnopar{\phi} \iff \overline{\astruct} \models \neg\phi \iff
  \beta_{\overline{\astruct}} \models \neg\bool{n}{\phi}$ again, by
  Lemma \ref{lemma:wss-bool} and the definition of $\bool{n}{\neg\phi}
  = \neg\bool{n}{\phi}$. Let $\overline{\beta}$ be the boolean
  valuation defined as $\overline{\beta}(\apred_k) =
  \neg\beta(\apred_k)$ for all $\apred\in\preds$ and $k\in[n]$. Then
  clearly $\overline{\beta} = \beta_{\overline{\astruct}}$ and
  $\overline{\beta} \models \neg\bool{n}{\phi} \iff \beta \models
  \dualnopar{\bool{n}{\phi}}$ follows. \qed}

\fi 

\subsection{Trap Invariants as Automata}
\label{sec:trap-automata}

The purpose of introducing automata is the definition of a
positivation operator for \ils\ or, equivalently, for \wss\ formulae.
Recall that, given a formula $\phi$, the result of positivation is a
formula $\ppos{\phi}$ in which all predicate symbols occur under an
even number of negations and, moreover $\phi \minequiv \ppos{\phi}$.

Unlike dualization, positivation is not defined on formulae but on
equivalent automata on finite words, obtained via the classical
two-way translation between \wss\ and Rabin-Scott automata, described
next. Let us fix a structure $\astruct = ([n],\nu,\iota,\mu)$ such
that $\dom(\nu) = \set{x_1,\ldots,x_k}$ , $\dom(\iota) =
\set{\apred_1, \ldots, \apred_\ell}$ and $\dom(\mu) =
\set{X_1,\ldots,X_m}$ are all finite.  Each such structure is viewed
as a word $w_\astruct = \sigma_0 \ldots \sigma_{n-1}$ of length $n$
over the alphabet $\set{0,1}^{k+\ell+m}$, where, for all $i \in [n]$,
we have: \begin{compactitem}
\item $\sigma_i(j)=1$ if $\nu(x_j)=i$ and $\sigma_i(j)=0$ otherwise,
  for all $1 \leq j \leq k$,
\item $\sigma_i(j)=1$ if $i \in \iota(\apred_{j-k})$ and
  $\sigma_i(j)=0$ otherwise, for all $k < j \leq k+\ell$,
\item $\sigma_i(j)=1$ if $i \in \mu(X_{j-k-\ell})$ and $\sigma_i(j)=0$
  otherwise, for all $k+\ell < j \leq k+\ell+m$.
\end{compactitem}
In other words, the $j$-th track of $w$ encodes \begin{inparaenum}[(i)]
\item the unique value $w(x_j) \isdef \nu(x_j)$, if $1 \leq j \leq k$,
\item the set $w(\apred_j) \isdef \iota(\apred_{j-k})$, if $k < j \leq
  k+\ell$, or
\item the set $w(X_j) \isdef \mu(X_{j-k-\ell})$, if $k+\ell < j \leq
  k+\ell+m$.
  \end{inparaenum}
For an alphabet symbol $\sigma \in \set{0,1}^{k+\ell+m}$, we write
$\sigma(\apred_j)$ for $\sigma(j+k)$.

\ifLongVersion
\begin{example}\label{ex:wss-word}
  \vspace*{-0.5\baselineskip}
  Consider the structure $\astruct=([6],\nu,\iota,\mu)$, where
  $\nu(x_1)=3$, $\iota(\apred_1)=\set{0,2,5}$ and
  $\mu(X_1)=\set{1,3}$. Moreover, assume that $\nu$, $\iota$ and $\mu$
  are undefined elsewhere. The word $w_\astruct$ is given below:
  \vspace*{-.5\baselineskip}
  \begin{center}
    \begin{tabular}{c|cccccc}
      & ~0~ & ~1~ & ~2~ & ~3~ & ~4~ & ~5 \\
      \hline
      $x_1$ & 0 & 0 & 0 & 1 & 0 & 0 \\
      $\apred_1$ & 1 & 0 & 1 & 0 & 0 & 1 \\
      $X_1$ & 0 & 1 & 0 & 1 & 0 & 0 
    \end{tabular}
  \end{center}
  \vspace*{-0.5\baselineskip}
  \hfill$\blacksquare$
\end{example}
\fi 

Given $x_1, \ldots, x_k \in \vars$, $\apred_1,\ldots,\apred_\ell \in
\preds$ and $X_1,\ldots,X_m \in \Vars$, a nondeterministic finite
automaton over the alphabet $\set{0,1}^{k+\ell+m}$ is a tuple $A =
(Q,I,F,\delta)$, where $Q$ is the finite set of states, $I \subseteq
Q$ is the set of initial states, $F \subseteq Q$ is the set of final
states and $\delta \subseteq Q \times \set{0,1}^{k+\ell+m} \times Q$
is the transition relation. A given a word $w=\sigma_0 \ldots
\sigma_{n-1}$ as before, a run of $A$ over $w$ is a sequence of states
$\rho = s_0 \ldots s_n$, such that $s_0 \in I$ and
$(s_i,\sigma_i,s_{i+1}) \in \delta$, for all $i \in [n]$. The run is
\emph{accepting} if $s_n \in F$, in which case we say that $A$ accepts
the word $w$. The language of $A$, denoted by $\lang{A}$, is the set
of words accepted by $A$. The following theorem is automata-theoretic
folklore\footnote{See e.g.\ \cite[Theorem 2.10.1 and
    2.10.3]{KhoussainovNerode}.}:

\begin{theorem}\label{thm:wss-automata}
  For each \wss\  formula $\phi(x_1, \ldots, x_k, \apred_1, \ldots,
  \apred_\ell, X_1, \ldots, X_m)$ there exists an automaton $A_\phi$
  over the alphabet $\set{0,1}^{k+\ell+m}$ such that $\astruct \models
  \phi \iff w_\astruct \in \lang{A}$, for each structure
  $\astruct$. Conversely, for each automaton $A$ over the alphabet
  $\set{0,1}^\ell$, there exists a \wss\  formula $\Phi_A(\apred_1,
  \ldots, \apred_\ell)$ such that $w_\astruct \in \lang{A} \iff
  \astruct \models \Phi_A$, for each structure
  $\astruct=([n],\nu,\iota,\mu)$ such that
  $\dom(\iota)=\set{\apred_1,\ldots,\apred_\ell}$ and
  $\dom(\nu)=\dom(\mu)=\emptyset$.
\end{theorem}
The construction of $A_\phi$ for the first point (logic to automata)
is by induction on the structure of $\phi$. The main consequence of
this construction is the decidability of the satisfiability problem
for the \wss\  logic, implied by the decidability of emptiness for
finite automata. Incidentally, this also proves the decidability of
\ils, as a consequence of Lemma \ref{lemma:ils-wss}. The second point
(automata to logic) is a bit less known and deserves presentation. Given
$A=(Q,I,F,\delta)$ with alphabet $\set{0,1}^\ell$ and states
$Q=\set{s_1,\ldots,s_q}$, we define a formula $\Psi_A(\apred_1,
\ldots, \apred_\ell, X_1,\ldots,X_q) \isdef \psi_{\mathit{cover}}
\wedge \psi_I \wedge \psi_\delta \wedge \psi_F$, where:
\[\begin{array}{rcl}
  \psi_{\mathit{cover}} & \isdef & \forall x ~.~ \bigvee_{i=1}^q X_i(x) \wedge \bigvee_{1\leq i < j \leq q} \neg X_i(x) \vee \neg X_j(x) \\
  \psi_I  & \isdef & \bigvee_{s_i \in I} X_i(\zero) \hspace*{4.8cm}
  \psi_F \isdef \exists x \forall y ~.~ y \leq x \wedge \bigvee_{s_i \in F} X_i(x) \\
  \psi_\delta & \isdef & \forall x \forall y ~.~ y \leq x \vee \bigvee_{(s_i,\sigma,s_j) \in \delta} X_i(x) \wedge X_j(\succ(x)) \wedge
  \bigwedge_{\!\!\!\!\!\!\begin{array}{c}
      \scriptstyle{1 \leq k \leq \ell} \\[-2mm]
      \scriptstyle{\sigma(\apred_k)=1}
  \end{array}} \!\!\!\!\apred_k(x) \wedge
  \bigwedge_{\!\!\!\!\!\!\begin{array}{c}
      \scriptstyle{1 \leq k \leq \ell} \\[-2mm]
      \scriptstyle{\sigma(\apred_k)=0}
  \end{array}} \!\!\!\!\neg\apred_k(x)
\end{array}\]
Intuitively, each $X_i$ keeps the positions that are labeled by the
state $s_i$ during the run of $A$ over some input word $w$ of length
$n$.  First, each position between $0$ and $n-1$ must be labeled with
exactly one state from $Q$ ($\psi_{\mathit{cover}}$). The initial
($\psi_I$) and final ($\psi_F$) positions are labeled with states from
$I$ and $F$, respectively. Next, each pair of adjacent positions is
labeled with a pair of states that is compatible with the transition
relation of $A$, on the corresponding input symbol, encoded as the
tuple $(w(\apred_1), \ldots, w(\apred_\ell)) \in \set{0,1}^\ell$
($\psi_\delta$). Finally, we define $\Phi_A \isdef \exists X_1 \ldots
X_q ~.~ \Psi_A$, to capture the fact that a word $w$ is accepted by
$A$ if and only if there exists an accepting run of $A$ over $w$.

Given a \wss\ formula $\phi$, we define a positivation operation
$\ppos{\phi}$ by translating first $\phi$ into an automaton
$A_\phi$. Then we \emph{saturate} $A_\phi$ by adding new transitions
to it, such that the language of the new automaton $A^*_\phi$ contains
$\lang{A_\phi}$ and the words corresponding to minimal structures are
the same in both $\lang{A_\phi}$ and $\lang{A^*_\phi}$. Then we obtain
$\ppos{\phi}$ by a slightly modified translation of $A^*_\phi$ into
\wss, which is guaranteed to produce positive formulae only. Note that
the result $\Phi_{A_\phi}$ of the above translation is not positive,
due to the formula $\psi_\delta$ which introduces negative predicates.

The saturation of an automaton $A = (Q,I,F,\delta)$ over the alphabet
$\set{0,1}^\ell$ is defined next. For each transition $(s,\sigma,s')
\in \delta$ the set $\delta^*$ contains all transitions $(s,\tau,s')$
such that $\tau \in \set{0,1}^\ell$ and $\sigma(j) \leq \tau(j)$, for
all $1 \leq j \leq \ell$. Moreover, nothing else is in $\delta^*$ and
$A^* \isdef (Q,I,F,\delta^*)$. In other words, $A^*$ is obtained by
adding to $A$, for each transition whose $j$-th track is $0$, another
transition in which this track is $1$.

To state the formal relation between $A$ and $A^*$, we define a
partial order on words over the alphabet $\set{0,1}^\ell$, encoding
the interpretations of the predicates $\apred_1, \ldots, \apred_\ell$:
$w_1 \wordeq w_2 \iff w_1(\apred_j) \subseteq w_2(\apred_j)$ for all
$1 \leq j \leq \ell$. The minimal language of $A$ is $\minlang{A}
\isdef \set{w \in \lang{A} \mid \forall w' ~.~ w' \wordeq w \wedge w'
  \neq w \Rightarrow w' \not\in \lang{A}}$.

\begin{lemma}\label{lemma:saturation}
  Given an automaton $A$ over the alphabet $\set{0,1}^\ell$, we have
  $\minlang{A} = \minlang{A^*}$. 
\end{lemma}
\proof{We start from the observation that $\lang{A} \subseteq
  \lang{A^*}$ because $A^* = (Q,I,F,\delta^*)$ is obtained by adding
  transitions to $A = (Q,I,F,\delta)$. Moreover, given a run $\rho =
  s_0, \ldots, s_m$ of $A^*$ over some word $\sigma_0 \ldots
  \sigma_{m-1}$, there exists a word $\sigma'_0 \ldots \sigma'_{m-1}$
  such that for each $i \in [m]$ and $1 \leq j \leq \ell$, we have
  $\sigma_i(j) \leq \sigma'_i(j)$ and $(s_i,\sigma'_i,s_{i+1}) \in
  \delta^*$ . This is because we only add to $A^*$ transitions
  $(q_i,\sigma',q_j)$ such that $\sigma(j) \leq \sigma'(i)$, for all
  $1 \leq j \leq \ell$, where $(q_i, \sigma, q_j) \in \delta$. 

  \noindent''$\subseteq$'' Let $w \in \minlang{A}$, then $w \in
  \lang{A^*}$ because $\lang{A} \subseteq \lang{A^*}$. Let $w'$ be a
  word such that $w' \wordeq w$ and $w' \neq w$ and suppose, for a
  contradiction that $w' \in \minlang{A^*}$. Then $A^*$ has an
  accepting run $\rho = s_0, \ldots, s_m$ over $w'$, thus $\rho$ is
  also an accepting run of $A$ over another word $w'' \wordeq
  w'$. Since $w \in \minlang{A}$ and $w'' \wordeq w' \wordeq w$ and
  $w' \neq w$, we obtain a contradiction. Thus, $w \in \minlang{A^*}$,
  as required.

  \noindent''$\supseteq$'' Let $w \in \minlang{A^*}$ and let
  $\rho=s_0, \ldots, s_m$ be an accepting run of $A^*$ over $w$. Then
  there exists a word $w' \wordeq w$ such that $\rho$ is an accepting
  run of $A$. Since $w' \in \lang{A} \subseteq \lang{A^*}$, we obtain
  that $w'=w$, thus $w \in \lang{A}$. Now suppose, for a
  contradiction, that there exists $w'' \wordeq w$ such that $w'' \neq
  w$ and $w'' \in \lang{A}$. Then $w'' \in \lang{A^*}$ and since $w''
  \wordeq w$ and $w'' \neq w$, this contradicts the fact that $w \in
  \minlang{A^*}$. Thus, $w \in \minlang{A}$, as required. \qed}

Finally, we define $\ppos{\phi} \isdef \exists X_1 \ldots X_q ~.~
\psi_{\mathit{cover}} \wedge \psi_I \wedge \psi^*_\delta \wedge
\psi_F$ as the formula obtained from $A_\phi=(Q,I,F,\delta)$ by
applying the translation scheme above in which, instead of
$\psi_\delta$, we use the following formula:
\[\begin{array}{c}
  \psi^*_\delta \isdef \forall x \forall y ~.~ y \leq x \vee
  \bigvee_{(s_i,\sigma,s_j) \in \delta} X_i(x) \wedge X_j(\succ(x)) \wedge
  \bigwedge_{\begin{array}{c}
      \scriptstyle{1 \leq k \leq \ell} \\[-2mm]
      \scriptstyle{\sigma(\apred_k)=1}
  \end{array}} \apred_k(x)
\end{array}\]
Note that $\ppos{\phi}$ is a positive formula, independently of
whether $\phi$ is positive or not. The following lemma proves the
required property of this positivation operation.

\begin{lemma}\label{lemma:positivation}
  Given a \wss\  sentence $\phi(\apred_1, \ldots, \apred_\ell)$, the
  following hold: \begin{compactenum}
  \item\label{it1:positivation} $\astruct \models \ppos{\phi} \iff
    w_\astruct \in \lang{A^*_\phi}$, for each structure
    $\astruct=([n],\nu,\iota,\mu)$ such that
    $\dom(\iota)=\set{\apred_1,\ldots,\apred_\ell}$ and
    $\dom(\nu)=\dom(\mu)=\emptyset$.
  \item\label{it2:positivation} $\phi \minequiv \ppos{\phi}$. 
  \end{compactenum}
\end{lemma}
\proof{(\ref{it1:positivation}) It is sufficient to show that
  $\ppos{\phi} \equiv \Phi_{A^*_\phi}$ and apply Theorem
  \ref{thm:wss-automata}. Denoting $A = (Q,I,F,\delta)$, with $Q =
  \set{s_1,\ldots,s_q}$ and $A^*=(Q,I,F,\delta^*)$ as before, we only
  show that $\psi^*_\delta \equiv \psi_{\delta^*}$. Because
  $\ppos{\phi} = \exists X_1 \ldots \exists X_q ~.~
  \psi_{\mathit{cover}} \wedge \psi_I \wedge \psi^*_\delta
  \wedge\psi_F$ and $\Phi_{A^*_\phi} = \exists X_1 \ldots \exists X_q
  ~.~ \psi_{\mathit{cover}} \wedge \psi_I \wedge \psi_{\delta^*}
  \wedge\psi_F$, we immediately obtain the result. We have the
  following equivalence, for each $\sigma \in \set{0,1}^\ell$:
  \[\bigwedge_{\begin{array}{c}
      \scriptstyle{1 \leq k \leq \ell} \\[-2mm]
      \scriptstyle{\sigma(\apred_k) = 1}
  \end{array}} \apred_k(x) \equiv
  \bigvee_{\sigma \leq \tau} \Big(\bigwedge_{\begin{array}{c}
      \scriptstyle{1 \leq k \leq \ell} \\[-2mm]
      \scriptstyle{\tau(\apred_k)=1} 
      \end{array}}\apred_k(x) \wedge \bigwedge_{\begin{array}{c}
      \scriptstyle{1 \leq k \leq \ell} \\[-2mm]
      \scriptstyle{\tau(\apred_k)=0} 
  \end{array}}\neg\apred_k(x)\Big)\]
  where $\sigma \leq \tau$ stands for $\forall j ~.~ 1 \leq j \leq
  \ell \Rightarrow \sigma(j) \leq \tau(j)$. This immediately implies
  that $\psi^*_\delta \equiv \psi_{\delta^*}$, by the definitions of
  these formulae and the construction of $\delta^*$.

  \vspace*{\baselineskip}
  \noindent(\ref{it2:positivation}) For an arbitrary structure
  $\astruct = ([n],\nu,\iota,\mu)$ we have $\iota(\apred_k) =
  w_\astruct(\apred_k)$, for any $1 \leq k \ell$, by the definition of
  $w_\astruct$. Then $\astruct_1 \strord \astruct_2 \iff
  w_{\astruct_1} \wordeq w_{\astruct_2}$, for any two structures
  $\astruct_i = ([n],\nu_i,\iota_i,\mu_i)$, where $i=1,2$. Hence a
  structure $\astruct$ is a minimal model of $\phi$ if and only if
  $w_\astruct \in \minlang{A_\phi}$. By Lemma \ref{lemma:saturation},
  we have $\minlang{A} = \minlang{A^*}$. Then the result follows from
  Theorem \ref{thm:wss-automata} and point (\ref{it1:positivation}) of
  this Lemma. \qed}

\ifLongVersion
Positivation and booleanization are related via the following property:

\begin{lemma}\label{lemma:pos-bool}
  Given a \wss\  formula $\phi$ and a constant $n>0$, we have
  $\pos{\bool{n}{\phi}} \equiv \bool{n}{\ppos{\phi}}$.
\end{lemma}
\proof{ First, note that, for any propositional formulae $f$ and $g$,
  whose variables occur under even number of negations, we have $f
  \equiv g \iff f \minequiv g$. Since both $\pos{\bool{n}{\phi}}$ and
  $\bool{n}{\ppos{\phi}}$ are positive propositional formulae, it is
  sufficient to prove $\pos{\bool{n}{\phi}} \minequiv
  \bool{n}{\ppos{\phi}}$, by showing $\minsem{\pos{\bool{n}{\phi}}}
  \subseteq \sem{\bool{n}{\ppos{\phi}}}$ and
  $\minsem{\bool{n}{\ppos{\phi}}} \subseteq
  \sem{\pos{\bool{n}{\phi}}}$, respectively, which establishes
  $\minsem{\pos{\bool{n}{\phi}}} = \minsem{\bool{n}{\ppos{\phi}}}$
  (the latter step is left to the reader).
  
  \noindent ``$\minsem{\pos{\bool{n}{\phi}}} \subseteq
  \sem{\bool{n}{\ppos{\phi}}}$'' Let $\beta \in
  \minsem{\pos{\bool{n}{\phi}}}$ be a valuation. Then, we also have
  $\beta \in \minsem{\bool{n}{\phi}}$, since $\pos{\varphi} \minequiv
  \varphi$, in general for any propositional formula $\varphi$. Then,
  by Lemma \ref{lemma:wss-bool}, there exists a structure $\astruct
  \in \minsem{\phi}$ such that $\beta = \beta_\astruct$. Hence we
  obtain $\astruct \in \minsem{\ppos{\phi}} \subseteq
  \sem{\ppos{\phi}}$. But then $\beta \in
  \sem{\bool{n}{\ppos{\phi}}}$, by Lemma \ref{lemma:wss-bool}.
  
  \noindent ``$\minsem{\bool{n}{\ppos{\phi}}} \subseteq
  \sem{\pos{\bool{n}{\phi}}}$'' Let $\beta \in
  \minsem{\bool{n}{\ppos{\phi}}}$ be a boolean valuation. By Lemma
  \ref{lemma:wss-bool}, we obtain a structure $\astruct \in
  \minsem{\ppos{\phi}}$ such that $\beta = \beta_\astruct$. But then
  $\astruct \in \minsem{\phi}$ and $\beta \in
  \minsem{\bool{n}{\phi}}$, by Lemma \ref{lemma:wss-bool}. Hence
  $\beta \in \sem{\pos{\bool{n}{\phi}}}$. \qed}
\fi
We are now ready to state the main result of the paper, concerning the
computation of trap invariants for parametric component-based systems.

\begin{theorem}\label{thm:param-trap-invariant}
  Given a parametric component-based system $\asys =
  \tuple{\typeno{\acomptype}{1}, \ldots,
    \typeno{\acomptype}{K},\interform}$, where $\typeno{\acomptype}{k}
  = \tuple{\typeno{\ports}{k}, \typeno{\states}{k},
    \typeno{\initstate}{k}, \typeno{\rules}{k}}$, for all
  $k=1,\ldots,K$, for any integer $n>0$ we have: 
  \[\alltrap{\amarkednet^n_\asys} \equiv
  \bool{n}{\dual{\ppos{\init{\asys} \wedge
        \mathit{Tr}\left(\trapconstraint{\interform}\right)}}}\] where
  $\init{\asys} \isdef \exists x ~.~ \bigvee_{k=1}^K
  \typeno{\initstate}{k}(x)$.
\end{theorem}
\proof{ Let $\amarkednet^n_\asys = (\anet,\amark_0)$ and $\mu_0 =
  \bigvee_{\amark_0(s)=1} s$. By Lemma \ref{lemma:pn-trap-invariant},
  we have \(\alltrap{\amarkednet_\asys} \equiv \dual{\pos{\mu_0 \wedge
      \trapconstraint{\anet}}}\).  From the definition of
  $\amarkednet_\asys$, it is not difficult to show that $\mu_0 \equiv
  \bool{n}{\init{\asys}}$ and $\trapconstraint{\anet} \equiv
  \bool{n}{\mathit{Tr}\left(\trapconstraint{\interform}\right)}$,
  hence $\mu_0 \wedge \trapconstraint{\anet} \equiv
  \bool{n}{\init{\asys} \wedge
    \mathit{Tr}\left(\trapconstraint{\interform}\right)}$. By Lemma
  \ref{lemma:pos-bool}, we obtain $\pos{\mu_0 \wedge
    \mathit{Tr}\left(\trapconstraint{\anet}\right)} \equiv
  \bool{n}{\ppos{\init{\asys} \wedge
      \mathit{Tr}\left(\trapconstraint{\interform}\right)}}$ and, by
  Lemma \ref{lemma:bool-dual}, we obtain $\dual{\pos{\mu_0 \wedge
      \trapconstraint{\anet}}} \equiv
  \bool{n}{\dual{\ppos{\init{\asys} \wedge
        \mathit{Tr}\left(\trapconstraint{\interform}\right)}}}$, as
  required. \qed} In practice, it is more efficient to perform
dualization directly on the saturated automaton $A^*_\phi$ for a given
\wss\ formula $\phi$ with predicate symbols $\apred_1, \ldots,
\apred_\ell$. To this end, we swap the $0$'s and $1$'s on the tracks
corresponding to $\apred_1, \ldots, \apred_\ell$ in the transition
rules of $A^*_\phi$ and complement the resulting automaton, call it
$\widetilde{A}_\phi$. Using Lemma \ref{lemma:wss-dual}, it is not
difficult to show that the complement of $\widetilde{A}_\phi$
corresponds to the formula $\dualnopar{\ppos{\phi}}$, needed to
compute the trap invariant of a system. A further optimization, that
avoids complementation of $\widetilde{A}_\phi$, is to check the
inclusion of the automaton $A_\psi$, obtained from the safety property
to be checked (i.e.\ $\psi$ may encode the states where a deadlock or
mutual exclusion violation occurs) into $\widetilde{A}_\phi$, using
state-of-the-art antichain or simulation-based inclusion checkers.
For this reasons, our experiments were carried out using the
\textsc{VATA} \cite{Vata} tree automata library as a decision
procedure for inclusion.

\paragraph{\em Remark~}
We argue that the trap invariant synthesis method given by Theorem
\ref{thm:param-trap-invariant} can be easily extended to handle
unbounded tree-like (hierarchical) systems. To this end, we consider a
variant of \ils\ equipped with a countably infinite set of successor
functions $\succ_0, \succ_1, \ldots$ ($\succ_0$ being the leftmost
successor) interpreted over the set $\nat^*$ of strings of natural
numbers, that identify positions in a tree as $\succ_k(k_0 \ldots k_m)
\isdef k_0 \ldots k_m k$. Also, the inequality is interpreted by the
prefix relation between strings. This logic is embedded into
\wssomega, the weak monadic second order logic of countably many
successors. Akin to the finite word case, \wssomega\ formulae can be
translated into (bottom-up nondeterministic) tree automata over finite
trees with symbolic (binary) alphabet, on which positivation and
dualization can be implemented similar to the word case. Moreover,
efficient antichain/simulation-based inclusion checks are also
available for tree automata \cite{Holik11}, thus expensive
complementation can be avoided in this case too. In
\S\ref{sec:experiments} we present an example involving a parametric
hierarchical tree architecture. A detailed workout of this
generalization is left for the future. \hfill$\blacksquare$

\section{Refining Invariants}

\begin{figure}[htb]
\vspace*{-\baselineskip}
\caption{Alternating Dining Philosophers}
\label{fig:philosophers2}
\vspace*{-\baselineskip}
\begin{center}
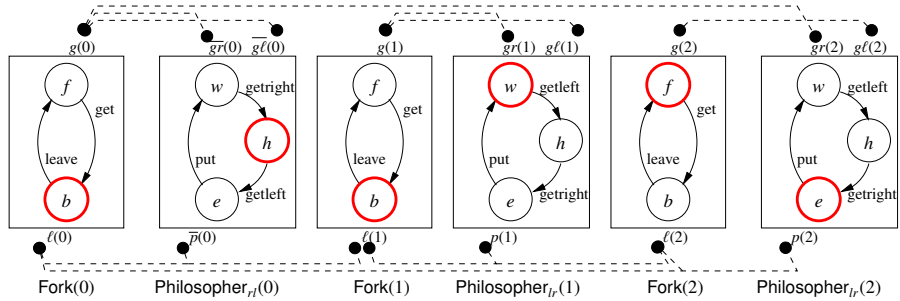
\end{center}
\vspace*{-2\baselineskip}
\end{figure}

Since the safety verification problem is undecidable for parametric
systems \cite{AptKozen86}, the trap invariants method cannot be
complete. As an example, consider the alternating dining philosophers
system, of which an instance (for $n=3$) is shown in
Fig. \ref{fig:philosophers2}. The system consists of two philosopher
component types, namely $\mathsf{Philosopher}_{rl}$, which takes its
right fork before its left fork, and $\mathsf{Philosopher}_{lr}$,
taking the left fork before the right one. Each philosopher has two
interaction ports for taking the forks, namely $g\ell$ (get left) and
$gr$ (get right) and one port for releasing the forks $p$ (put). The
ports of the $\mathsf{Philosopher}_{rl}$ component type are overlined,
in order to be distinguished. The $\mathsf{Fork}$ component type is
the same as in Fig. \ref{fig:philosophers}. The interaction formula
for this system $\interform_{\mathit{philo}}^{\mathit{alt}}$, shown in
Fig. \ref{fig:philosophers2}, implicitly states that only the
$0$-index philosopher component is of type
$\mathsf{Philosopher}_{rl}$, whereas all other philosophers are of
type $\mathsf{Philosopher}_{lr}$. Note that the interactions on ports
$\overline{g\ell}$, $\overline{gr}$ and $\overline{p}$ are only
allowed if $\mathit{inf}(x)$ holds, i.e.\ $x=0$.

It is well-known that any instance of the parametric alternating
dining philosophers system consisting of at least one
$\mathsf{Philosopher}_{rl}$ and one $\mathsf{Philosopher}_{lr}$ is
deadlock-free. However, trap invariants are not enough to prove
deadlock freedom, as shown by the global state $\set{b(0), h(0), b(1),
  w(1), f(2), e(2)}$, marked with thick red lines in Fig.
\ref{fig:philosophers2}. Note that no interaction is enabled in this
state. Moreover, this state intersects with any trap of the marked PN
that defines the executions of this particular instance, as proved
below. Consequently, the trap invariant contains a deadlock
configuration, and the system cannot be proved deadlock-free by this
method.

\begin{proposition}\label{prop:cex}
  Consider an instance of the alternating dining philosophers system
  in Fig. \ref{fig:philosophers2}, consisting of components
  $\mathsf{Fork}(0)$, $\mathsf{Philosopher}_{rl}(0)$,
  $\mathsf{Fork}(1)$, $\mathsf{Philosopher}_{lr}(1)$,
  $\mathsf{Fork}(2)$ and $\mathsf{Philosopher}_{lr}(2)$ placed in a
  ring, in this order. Then each nonempty trap of this system contains
  one of the places $(b,0), (h,0), (b,1), (w,1), (f,2)$ or $(e,2)$.
\end{proposition}
\proof{ Let $C = \set{b(0), h(0), b(1), w(1), f(2), e(2)}$ in the
  following. We shall try to build a nonempty trap $T$ that avoids
  every state in $C$. If such a trap can be found, the counterexample
  is shown to be spurious (unreachable). Below is the list of states
  allowed in $T$, indexed by component (using other states that the
  ones listed below would result in a trap that is satisfied by the
  counterexample $C$, which is exactly the opposite of what we want):

\[\begin{array}{c|c|c|c|c|c}
\mathsf{Fork}(0) & \mathsf{Philosopher}_{rl}(0) & \mathsf{Fork}(1) & \mathsf{Philosopher}_{lr}(1) & \mathsf{Fork}(2) & \mathsf{Philosopher}_{lr}(2) \\
\hline
f(0) & w(0), e(0) & f(1) & h(1), e(1) & b(2) & w(2), h(2)
\end{array}\]

Assume that $f(0) \in T$. Then $T$ must contain $b(0)$ or $e(2)$
(constraint $gr(2) \wedge g(0)$). However neither is allowed, thus
$f(0) \not\in T$. Assume that $f(1) \in T$. Then $T$ must contain
$b(1)$ or $h(0)$ (constraint $gr(0) \wedge g(1)$), contradiction, thus
$f(1) \not\in T$. Assume that $b(2) \in T$. Then $T$ must contain
$f(1), w(1)$ or $f(2)$ (constraint $p(1) \wedge \ell(1) \wedge
\ell(2)$), contradiction, thus $b(2) \not\in T$. Then $T$ contains
only philosopher states, except for $h(0)$, $w(1)$ and $e(2)$. One can
prove that there is no such trap, for instance, for
$\mathsf{Philosopher}_{lr}(1)$ we have:
\[\begin{array}{rcl}
h(1) \in T & \Rightarrow & e(1) \in T \\
e(1) \in T & \Rightarrow & w(1) \in T
\end{array}\]
since $f(1), b(1), f(2), b(2) \not\in T$. Since $w(1) \not\in T$, we
obtain that $h(1), e(1) \not\in T$. Then the only possibility is
$T=\emptyset$. \qed}

However, the configuration is unreachable by a real execution of the
PN, started in the initial configuration $\bigwedge_{i=0}^2 f(i)
\wedge w(i)$. An intutive reason is that, in any reachable
configuration, each fork is in state $f$(ree) only if none of its
neighbouring philosophers is in state $e$(ating). In order to prove
deadlock freedom, one must learn this and other similar
constraints. Next, we present a heuristic method for strenghtening the
trap invariant, that learns such universal constraints, involving a
fixed set of components.

\subsection{Ashcroft Invariants}

\emph{Ashcroft invariants} (AI) \cite{Ashcroft75} are a classical
method for proving safety properties of parallel programs, in which a
global program state is viewed as an array consisting of the local
states of each thread. Typically, an AI is an universally quantified
assertion $\forall x_1 \ldots x_m ~.~ \phi$ that relates the local
states of at most $m$ distinct threads, which is, moreover, an
invariant of the execution of the parallel program.

Next, we define a variant of AI tuned for our purposes. We first
consider a finite \emph{window} of the parametric system, by
identifying a fixed set of components, together with their
interactions, and abstracting away all interactions among components
outside of this window. The crux is that the indices of the components
from the window are not numbers but Skolem constants $c_1, \ldots,
c_w$ and the window is defined by a logical formula $\psi(c_1, \ldots,
c_w)$ over the vocabulary of these constants, involving inequalities
between terms of the form $s^i(c_j)$, for some $i \geq 0$. This allows
to slide the window inside a certain range, without changing
the view (i.e.\ the sub-systems observed by sliding the window are all
isomorphic). Each view is a finite component-based system, whose set
of reachable states is computable by enumerating the (finite set of)
reachable markings of a 1-safe PN of fixed size. Let
$\Phi(c_1,\ldots,c_w)$ be the formula defining this set. Then we show
that $\forall x_1 \ldots \forall x_w ~.~ \psi(x_1, \ldots, x_w)
\rightarrow \Phi(x_1,\ldots,x_w)$ defines an AI of the parametric
system $\asys$, that can be used to strenghten the trap invariant and
converge towards a proof of the given safety property. Before entering
the formal details, we provide an example.

\begin{figure}[htb]
  \vspace*{-\baselineskip}
  \caption{Window Petri Net for the Alternating Dining Philosophers Example}
  \label{fig:window}
  \vspace*{-\baselineskip}
  \begin{center}
    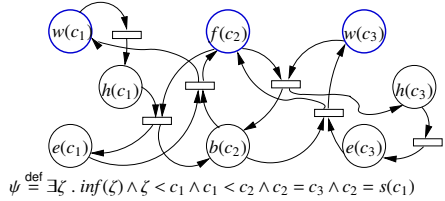
  \end{center}
  \vspace*{-2\baselineskip}
\end{figure}

\begin{example}\label{ex:philo-ai}
   Consider the alternating dining philosophers system, with
   interaction formula $\interform_{\mathit{philo}}^{\mathit{alt}}$,
   given in Fig. \ref{fig:philosophers2}. We fix three adjacent
   components, namely $\mathsf{Philosopher}_{lr}(c_1)$,
   $\mathsf{Fork}(c_2)$ and $\mathsf{Philosopher}_{lr}(c_3)$, such
   that the window constraint $\psi(c_1,c_2,c_3) \isdef \exists \zeta
   ~.~ \mathit{inf}(\zeta)$ $\wedge \zeta < c_1 \wedge c_1 < c_2 \wedge
   c_2 = \succ(c_1) \wedge c_2 = c_3$ holds. The interactions
   specified by $\interform_{\mathit{philo}}^{\mathit{alt}}$,
   involving nothing but these components are $gr(c_1) \wedge g(c_2)$,
   $p(c_1) \wedge \ell(c_2) \wedge p(c_3)$ and $g\ell(c_3) \wedge
   g(c_2)$. In addition, $\mathsf{Philosopher}_{lr}(c_1)$ interacts
   with its left fork, not present in the window defined by $\psi$. We
   abstract this interaction to $g\ell(c_1)$. The other partial
   interactions are $gr(c_3)$ and $p(c_3) \wedge \ell(c_2)$, where the
   fork to the right of $\mathsf{Philosopher}_{lr}(c_3)$ is missing
   from the window. The marked PN corresponding to the window is given
   in Fig. \ref{fig:window}, with the initial marking highlighted in
   blue. Let $\Phi(c_1, c_2, c_3)$ be the ground formula describing
   the set of reachable markings of this PN. The AI corresponding to
   this window is $\forall x_1 \forall x_2 \forall x_3 ~.~ \exists
   \zeta ~.~ \mathit{inf}(\zeta) \wedge \zeta < x_1 \wedge x_1 < x_2
   \wedge x_2 = x_3 \wedge x_2 = \succ(x_1) \rightarrow
   \Phi(x_1,x_2,x_3)$. In particular, this invariant excludes the
   spurious deadlock counterexample of Fig. \ref{fig:philosophers2} by
   ensuring that a fork is in state $f$(ree) only if none of its
   neighbouring $\mathsf{Philosopher}_{lr}$'s is in state
   $e$(ating). \hfill$\blacksquare$  
\end{example}

Let $\asys = \tuple{\typeno{\acomptype}{1}, \ldots,
  \typeno{\acomptype}{K},\interform}$ be a parametric component-based
system with component types $\typeno{\acomptype}{k} =
\tuple{\typeno{\ports}{k}, \typeno{\states}{k},
  \typeno{\initstate}{k}, \typeno{\rules}{k}}$, for all
$k=1,\ldots,K$ and an existential interaction formula:
\begin{equation}\label{eq:existential-interform}
\begin{array}{c}
\interform = \exists x_1 \ldots \exists x_m
\bigvee_{i=1}^\ell ~.~ \varphi_i(x_1, \ldots, x_m) \wedge
\bigwedge_{j=1}^{k_i} p_{ij}(x_{ij})
\end{array}
\end{equation}
where $\varphi_i$ is a quantifier-free \ils~ formula not involving
predicate atoms and $x_{ij} \in \set{x_1, \ldots, x_m}$, for all $i
\in \set{1, \ldots, \ell}$ and all $j \in \set{1, \ldots, k_i}$. For
example, the interaction formulae $\interform_{\mathit{philo}}$
(Fig. \ref{fig:philosophers}) and
$\interform_{\mathit{philo}}^{\mathit{alt}}$
(Fig. \ref{fig:philosophers2}) are both inside this class. In order to
define the notion of a window, we fix a set of constant symbols
$\vec{c}=\set{c_1, \ldots, c_w}$, each having an associated component
type, denoted by $\typeof{c_i} \in \set{\typeno{\acomptype}{1},
  \ldots, \typeno{\acomptype}{K}}$, for all $i = 1, \ldots, w$. Note
that we overload the $\typeof{.}$ notation to handle both predicate
and constant symbol arguments. For instance, in Example
\ref{ex:philo-ai}, we have $\typeof{c_1} = \typeof{c_3} =
\mathsf{Philosopher}_{lr}$ and $\typeof{c_2} = \mathsf{Fork}$.

\begin{definition}\label{def:non-overlapping}
  Given two \ils~ formulae $\phi_1$ and $\phi_2$, we write $\phi_1
  \models \phi_2$ for $\sem{\phi_1} \subseteq \sem{\phi_2}$.  Then
  $\phi_1$ is \emph{non-overlapping} with $\phi_2$ if and only if
  either $\phi_1 \models \phi_2$ or $\phi_1 \models \neg\phi_2$
  holds.
\end{definition}

A \emph{window constraint} for the interaction formula $\interform$ as
before (\ref{eq:existential-interform}), is a formula $\psi$ that is
non-overlapping with each of the formulae:
\[\begin{array}{c}
\phi_i(c_{i_1}, \ldots, c_{i_k}) \isdef \exists x_1\ldots \exists x_p ~.~
\bigwedge_{i=1}^{p} \bigwedge_{u=1}^w x_i \neq c_u \wedge
\varphi_i(y_1, \ldots, y_m)
\end{array}\]
where $0 \leq p \leq w$ is an integer, $\set{c_{i_1}, \ldots, c_{i_k}}
\subseteq \set{c_1, \ldots, c_w}$ and $\set{y_1,\ldots,y_m}$ is a
reindexing of the set $\set{x_1,\ldots,x_p} \cup \set{c_{i_1}, \ldots,
  c_{i_k}}$. Since there are finitely many such formulae\footnote{The
  set $\set{\phi_i(c_{i_1}, \ldots, c_{i_k}) \mid 1 \leq i \leq \ell,~
    c_{i_1}, \ldots, c_{i_k} \in \vec{c}}$ is determined by
  $\interform$ and $\vec{c}$.}, it is possible to build window
constraints, by taking conjunctions in which each $\phi_i(c_{i_1},
\ldots, c_{i_k})$ formula occurs either positively or under negation.

\ifLongVersion
\begin{definition}\label{def:isomorphism}
Given a set of constant symbols $\vec{c}$, two \ils-structures $([n],
\iota_1, \nu_1)$ and $([n], \iota_2, \nu_2)$ are
$\vec{c}$-\emph{isomorphic}, denoted $([n], \iota_1, \nu_1)
\approx_{\vec{c}} ([n], \iota_2, \nu_2)$ if and only if $\iota_1(c)
\in \iota_1(\apred) \iff \iota_2(c) \in \iota_2(\apred)$ for all $c
\in \vec{c}$ and all $\apred \in \preds$. For a structure $\I$, we
denote by $[\I]_{\vec{c}}$ its $\approx_{\vec{c}}$-equivalence class.
\end{definition}
\fi

\begin{definition}\label{def:view}
Given a window constraint $\psi$, the \emph{view} of $\interform$ via
$\psi$ is the ground formula \(\viewform{\interform}{\psi} \isdef
\bigvee_{i=1}^\ell \bigwedge_{j=1}^{k_i} \pi_{ij}\), where, for each
$1 \leq i \leq \ell$ and each $1 \leq j \leq k_i$, we have $\pi_{ij}
\isdef p_{ij}(c_{ij})$ if $\psi \models \phi_i(c_{i_1}, \ldots,
c_{i_k})$, for some $c_{i_1}, \ldots, c_{i_k} \in \vec{c}$ and
$\pi_{ij} \isdef \top$, otherwise. 
\end{definition}
Intuitively, the view specifies the complete interactions between the
components identified by $c_1, \ldots, c_w$ and their respective types
as well as all the partial interactions from which some component is
missing from the window, i.e.\ when $\set{c_{i_1}, \ldots, c_{i_k}}
\neq \vec{c}$. Note that each interaction is unambiguously specified
by the window constraint, because either \begin{inparaenum}[(i)]
  \item $p_{ij}(c_{ij})$ is part of the interaction then $\psi \models
    \phi_i$, thus the component identified by $c_{ij}$ and the
    component type $\typeof{c_{ij}}$ is always in the window (no
    matter what value does $c_{ij}$ take), or
  \item $\psi \not\models \phi_i$, and since $\psi$ is non-overlapping
    with $\phi_i$, we have $\psi \models \neg\phi_i$, in which case
    the $c_{ij}$ component is never within the $\psi$ window.
\end{inparaenum}

\begin{example}\label{ex:philo-ai-window}
  \vspace*{-0.5\baselineskip} For the system in
  Fig. \ref{fig:philosophers2} and the window constraint $\psi$ from
  Example \ref{ex:philo-ai}, we obtain the view
  $\viewform{\interform}{\psi} = (gr(c_1) \wedge g(c_2)) \vee (p(c_1)
  \wedge \ell(c_2) \wedge p(c_3)) \vee (g\ell(c_3) \wedge g(c_2)) \vee
  g\ell(c_1) \vee gr(c_3) \vee (p(c_3) \wedge \ell(c_2))$. The
  interaction $gr(c_1) \wedge g(c_2)$ occurs between components inside
  the window only, since $\psi \models \neg\inf(c_1) \wedge c_2 =
  \succ(c_1)$. On the other hand, $p(c_3) \wedge \ell(c_2)$ is a partial
  interaction, because $\psi \models \exists z_1 ~.~ c_2 = c_3 \wedge
  z_1 = \succ(c_3) \wedge p(c_3) \wedge \ell(c_2) \wedge \ell(z_1)$, thus
  ports $p(c_3)$ and $\ell(c_2)$ are kept inside and $\ell(\succ(c_3))$ is
  abstracted away. \hfill$\blacksquare$
  \vspace*{-0.5\baselineskip}
\end{example}

A view $\viewform{\interform}{\psi}$ becomes the interaction formula
of a system with a constant number of components, whose marked PN is
denoted by $\amarkednet_\asys^\psi$.
\ifLongVersion
Formally, we define
$\amarkednet_\asys^\psi = (\anet^\psi, \amark_0^\psi)$, where
$\anet^\psi = (\places^\psi, \trans^\psi, \edges^\psi)$
and: \begin{compactitem}
\item $\places^\psi \isdef \set{\succ(c) \mid s \in
  \states(\typeof{c}),~ c \in \vec{c}}$,
\item for each equivalence class $[\I]_{\vec{c}}$ of some
  $\I=([n],\iota,\nu) \in \sem{\viewform{\interform}{\psi}}$, there
  exists $\atrans \in \trans^\psi$ and $(\succ(c),\atrans),
  (\atrans,s'(c)) \in \edges^\psi$ iff $s \arrow{p}{} s' \in
  \rules(\typeof{c})$ and $\iota(c) \in \iota(p)$, for all $\succ(c) \in
  \places^\psi$,
\item for all $\succ(c) \in \places^\psi$, we have $\amark_0(\succ(c))=1$ iff
  $s = \initstate(\typeof{c})$.
\end{compactitem}
\fi
Since this is a $1$-safe marked PN of known size, it is possible to
compute its reachable markings by exhaustive enumeration and compute a
ground formula $\reachform{\interform}{\psi}(c_1, \ldots, c_w)$ that
defines this set. For instance, the marked PN for the view of the
system in Fig. \ref{fig:philosophers2} via the window constraint
$\psi$ from Example \ref{ex:philo-ai}, is given in
Fig. \ref{fig:window}.
\ifLongVersion
Its set of reachable markings is defined by the
formula\footnote{We intentionally left out the negative literals, as
  they play no role in proving deadlock freedom.}:
\[\begin{array}{c}
(w(c1) \wedge f(c2) \wedge w(c3)) \vee
(h(c1) \wedge f(c2) \wedge w(c3)) \vee
(w(c1) \wedge b(c2) \wedge h(c3)) \\ 
\vee~ (e(c1) \wedge b(c2) \wedge w(c3))
\vee (h(c1) \wedge b(c2) \wedge h(c3)) \\
\vee~ (w(c1) \wedge b(c2) \wedge e(c3))
\vee (h(c1) \wedge b(c2) \wedge e(c3))
\end{array}\]
\fi
Finally, this formula is used to build the AI $\forall x_1 \ldots
\forall x_w ~.~ \psi(x_1, \ldots, x_w) \rightarrow
\reachform{\interform}{\psi}(x_1, \ldots, x_w)$, where $\psi(x_1,
\ldots, x_w)$ and $\reachform{\interform}{\psi}(x_1, \ldots,x_w)$ are
obtained from $\psi$ and $\reachform{\interform}{\psi}$, respectively,
by replacing each constant symbol $c_i$ with a variable $x_i$. The
following result states that this formula defines an invariant of the
system:

\begin{proposition}\label{prop:ai}
  Let $\asys = \tuple{\typeno{\acomptype}{1}, \ldots,
    \typeno{\acomptype}{K},\interform}$ be a system and let $\psi(c_1,
  \ldots, c_w)$ be a window constraint. Then $\ainv{\interform}{\psi,n}
  \isdef \nsem{\forall x_1 \ldots \forall x_w ~.~ \psi(x_1, \ldots,
    x_w) \rightarrow \reachform{\interform}{\psi}(x_1,\ldots,x_w)}$ is
  an invariant of $\amarkednet_\asys^n$, for any $n>0$, where
  $\nsem{\phi} \isdef \set{([n], \iota, \nu) \mid ([n], \iota, \nu) \models
    \phi}$ for any \ils~ formula $\phi$.
\end{proposition}
\proof{ Let $n > 0$ be an arbitrary positive integer. The component
  types of $\asys$ are $\typeno{\acomptype}{k} =
  \tuple{\typeno{\ports}{k}, \typeno{\states}{k},
    \typeno{\initstate}{k}, \typeno{\rules}{k}}$, for all
  $k=1,\ldots,K$ and its marked PN $\amarkednet_\asys^n = (\anet,
  \amark_0)$, where: \begin{compactitem}
  \item $\anet = (\bigcup_{k=1}^K \typeno{\states}{k} \times [n],
    \trans, \edges)$ and,
  \item for all $1 \leq k \leq K$ and all $s \in \typeno{\states}{k}$,
    we have $\amark_0((s, i)) = 1$, if $s = \typeno{\initstate}{k}$
    and $\amark_0((s, i)) = 0$, otherwise.
  \end{compactitem}
  Moreover, for a marking $\amark : \bigcup_{k=1}^K
  \typeno{\states}{k} \times [n] \rightarrow \set{0,1}$ of
  $\amarkednet_\asys^n$ and a formula $\phi$, we write $\amark \in
  \nsem{\phi}$ iff $([n], \iota, \nu) \models \phi$, where $\iota$ is
  such that $\iota(s) = \set{i \in [n] \mid \amark((s,i)) = 1}$ and
  $\nu$ is an arbitrary valuation.
  
  We prove first that $\amark_0 \in \ainv{\interform}{\psi,n}$. Let
  $i_1, \ldots, i_w \in [n]$ be arbitrary integers. Clearly,
  $\amark_0((\typeno{\initstate}{k_1}, i_1)) = \ldots =
  \amark_0((\typeno{\initstate}{k_w}, i_w)) = 1$, thus $([n],
  \iota[\typeno{\initstate}{k_1} \leftarrow \set{i_1}] \ldots
       [\typeno{\initstate}{k_w} \leftarrow \set{i_w}], \nu[x_1
         \leftarrow i_1] \ldots [x_w \leftarrow i_w]) \models
       \reachform{\interform}{\psi}$, since
       $\reachform{\interform}{\psi}$ is the set of reachable markings
       of $\amarkednet_\asys^\psi$ and $\amark_0$ subsumes the initial
       marking thereof.

  Second, we show that $\ainv{\interform}{\psi,n}$ is inductive,
  i.e.\ for each move $\amark \arrow{\atrans}{} \amark'$ of
  $\amarkednet_\asys^n$, such that $\amark \in
  \ainv{\interform}{\psi,n}$, we must show that $\amark' \in
  \ainv{\interform}{\psi,n}$. First, notice that, for any
  interpretation $\iota$, any valuation $\nu$ and any $i_1, \ldots,
  i_w \in [n]$, we have:
  \begin{eqnarray}
    & ([n], \iota[c_1 \leftarrow i_1] \ldots [c_w \leftarrow i_w], \nu) \models \phi(c_1, \ldots, c_w) \nonumber & \\
    & \iff & \label{eq:symb} \\
    & ([n], \iota, \nu[x_1 \leftarrow i_1] \ldots [x_w \leftarrow i_w]) \models \phi(x_1, \ldots, x_w) \nonumber &
  \end{eqnarray}
  for an arbitrary formula $\phi$. In the following, we define, for
  all $s \in \bigcup_{k=1}^K \typeno{\states}{k}$:
  \[\begin{array}{rcl}
  \iota_\amark(s) & \isdef & \set{i \in [n] \mid \amark((s,i)) = 1} \\
  \iota_{\amark'}(s) & \isdef & \set{i \in [n] \mid \amark'((s,i)) = 1}
  \end{array}\]  
  Let $i_1, \ldots, i_w \in [n]$ be integers such that $([n],
  \iota_{\amark'}, \nu[x_1 \leftarrow i_1] \ldots [x_w \leftarrow
    i_w]) \models \psi(x_1, \ldots, x_w)$. We compute as follows:
  \[\begin{array}{rclll}
  ([n], \iota_{\amark'}, \nu[x_1 \leftarrow i_1] \ldots [x_w \leftarrow i_w]) & \models & \psi(x_1, \ldots, x_w) & \iff &
  \text{since $\psi$ has only atoms $s^k(x_i) \leq s^\ell(x_j)$} \\
  ([n], \iota_{\amark}, \nu[x_1 \leftarrow i_1] \ldots [x_w \leftarrow i_w]) & \models & \psi(x_1, \ldots, x_w) & \iff &
  \text{by (\ref{eq:symb})} \\
  ([n], \iota_{\amark}[c_1 \leftarrow i_1] \ldots [c_w \leftarrow i_w], \nu) & \models & \psi(c_1, \ldots, c_w) & \Longrightarrow &
  \text{since $\amark \in \ainv{\interform}{\psi,n}$} \\
  ([n], \iota_{\amark}[c_1 \leftarrow i_1] \ldots [c_w \leftarrow i_w], \nu) & \models & \reachform{\interform}{\psi}(c_1, \ldots, c_w) & \Longrightarrow &
  (\dagger) \\
  ([n], \iota_{\amark'}[c_1 \leftarrow i_1] \ldots [c_w \leftarrow i_w], \nu) & \models & \reachform{\interform}{\psi}(c_1, \ldots, c_w) & \iff &
  \text{by (\ref{eq:symb})} \\
  ([n], \iota_{\amark'}, \nu[x_1 \leftarrow i_1] \ldots [x_w \leftarrow i_w]) & \models & \reachform{\interform}{\psi}(x_1, \ldots, x_w) &&
  \text{thus $\amark' \in \ainv{\interform}{\psi,n}$, as required.}
  \end{array}\]
  We are left with proving the step ($\dagger$) above. Because
  $\amarkednet_\asys^n$ is $\asys$-decomposable, by Lemma
  \ref{lemma:decomposable-pn}, and since $\amark \arrow{\atrans}{}
  \amark'$ by the hypothesis, there are states $s_1, s'_1 \in
  \typeno{\states}{k_1}$ $\ldots$ $s_m, s'_m \in
  \typeno{\states}{k_m}$, with $k_1, \ldots, k_m$ pairwise disjoints,
  integers $j_1, \ldots, j_m \in [n]$ and edges $((s_i, j_i),
  \atrans), (\atrans, (s'_i, j_i)) \in \edges$, for all $i=1, \ldots,
  m$. For each $i = 1, \ldots, m$, we distinguish the cases: \begin{compactitem}
  \item if $s_i = s'_i$ then $\amark((s_i, j_i)) = \amark'((s_i, j_i)) = 1$, 
  \item else, if $s_i \neq s'_i$ then $\amark((s_i, j_i)) =
    \amark'((s'_i, j_i)) = 1$ and $\amark((s'_i, j_i)) = \amark'((s_i,
    j_i)) = 0$.
  \end{compactitem}
  Observe now that $\amarkednet_\asys^\psi$ has the same structure as
  the subnet obtained by restricting $\amarkednet_\asys^n$ to the
  states in $\set{s_1, \ldots, s_m, s'_1, \ldots, s'_m} \times
  \set{j_1, \ldots, j_m}$. Moreover, there exists a transition
  $\proj{\atrans}{\psi}$ in $\amarkednet_\asys^\psi$ and edges
  $(s_i(c_i), \proj{\atrans}{\psi)}, (\proj{\atrans}{\psi}, s'_i(c_i))
  \in E^\psi$ only if $((s_i,j_i), \atrans), (\atrans, (s'_i,j_i)) \in
  E$. Let $\proj{\amark}{\psi}$ and $\proj{\amark'}{\psi}$ be the
  projections of $\amark$ and $\amark'$ on $\set{s_1, \ldots, s_m,
    s'_1, \ldots, s'_m} \times \set{j_1, \ldots, j_m}$,
  respectively. Since $\amark \arrow{\atrans}{} \amark'$, we obtain
  that $\proj{\amark}{\psi} \arrow{\proj{\atrans}{\psi}}{}
  \proj{\amark'}{\psi}$, thus $([n], \iota_{\amark'}[c_1 \leftarrow
    i_1] \ldots [c_w \leftarrow i_w], \nu) \models
  \reachform{\interform}{\psi}(c_1, \ldots, c_w)$, as required. \qed}

\section{Experiments}
\label{sec:experiments}

We carried out a preliminary evaluation of our parametric verification
method, using a number of textbook examples, shown in Table
\ref{tab:experiments}. The table reports the size of the example
(number of states/transition per component type) and the running times
(in seconds) needed to check deadlock freedom (D-freedom) and mutual
exclusion (Mutex). We used the \textsc{MONA} tool \cite{Mona} to
generate the automata from \wss\ formulae and the \textsc{VATA} tree
automata library \cite{Vata} to check the verification condition on
automata. The running times from the table are relative to a
x86\_64bit Ubuntu virtual machine with 4GB or RAM. The files needed to
reproduce the results are available
online\footnote{\url{http://nts.imag.fr/images/0/06/Cav19.tar.gz}}. All
examples were successfully verified for deadlock freedom by our method
using trap invariants. However, not all experiments with mutual
exclusion were conclusive, as the intersection of the invariant with
the bad states was not empty in some cases. The dash from the Mutex
column indicates that mutual exclusion checking is not applicable for
the considered example.

\begin{table}
\vspace*{-\baselineskip}
\begin{center}
\begin{tabular}{|l|c|c|c|}
\hline
Example & States/Transitions & D-freedom (sec) & Mutex (sec) \\ 
\hline
Dining Philosophers I & 3/3 + 2/2 & 0.252 & $-$ \\
Dining Philosophers II  & 3/3 + 3/4 & 0.496 & $-$ \\
Dining Philosophers III & 3/3 + 2/2 & 21.640 & $-$ \\
Exclusive Tasks & 2/3 & 0.004 & 0.004 \\
Preemptive Tasks I & 4/5 & 0.020 & 1.612 \\
Preemptive Tasks II & 4/5 & 0.020 & 1.564 \\
Burns  & 6/8 & 0.012 & 0.012 \\
Szymanski & 12/13 & 2.892 & not empty \\
Dijkstra-Scholten I & 4/4 & 0.012 & $-$ \\
Dijkstra-Scholten II & 4/4 & 0.064 & $-$ \\ 
\hline
\end{tabular}
\caption{Running times deadlock-freedom and mutual exclusion checking}
\label{tab:experiments}
\end{center}
\vspace*{-3\baselineskip}
\end{table}

Dining Philosphers I is the alternating dinning philosophers protocol
where all but one philosopher are taking the forks in the same
order. This example requires an additional Ashcroft invariant for
deadlock freedom. Dining Philosphers II is a refinement of the
previous model, where the behavior of the forks remembers which
philosopher is handling them (using two busy states
$b_{\mathit{left}}$ and $b_{\mathit{right}}$). Dining Philosophers III
is a variant of the protocol, where the philosophers are sharing
two global forks, taken in the same order. In these two cases, the
trap invariant is sufficient to prove deadlock freedom.

Exclusive Tasks is a mutual exclusion protocol in which every task can
be waiting or executing. A task moves from waiting to executing only
if all other tasks are waiting, whereas an executing task can move
back from execution to waiting at any time. Preemptive Tasks I is a
concurrent system in which every task can be ready, waiting, executing
or preempting.  Initially, one task is executing, while the others are
ready. A task moves from ready to waiting at any time.  A task begins
execution by preempting the currently executing tasks. When a task
finishes it becomes ready and one the preempted tasks resumes back to
execution. Preemptive Tasks II is same as before, except that the task
which resumes back to execution is always the one with the highest
identifier.

Burns \cite{Jensen98} and Szymanski \cite{Szymanski90} are classical
mutual exclusion protocols taken from literature. Dijsktra-Scholten I
is an algorithm used to detect termination in a distributed
computation. It organizes the computational nodes into a tree and
propagates a message from the root to all the leaves and back, once
the computation is finished. In the first variant, we consider the
degenerate case where the tree is a list. Dijkstra-Scholten II is the
full version of the algorithm on an arbitrary binary tree. This
example required using \textsc{MONA} and \textsc{VATA} in tree mode,
on \wssomega\ and finite nondeterministic bottom-up tree automata,
respectively.

\enlargethispage{10mm}

\section{Conclusions and Future Work}

We presented a method for checking safety properties of parametric
systems, in which the number of components is not known a~priori. The
method is based on a synthesis of trap invariants from the interaction
formula of the system and relies on two logical operations
(positivation and dualization) that are implemented using the
automata-theoretic connection between \wss and finite Rabin-Scott
automata. We show that trap invariants, strenghtened with Ashcroft
invariants, produced by an orthogonal method are, in general, strong
enough to prove deadlock freedom.

As future work, we plan on developing a toolbox integrating the
existing tools used to generate trap and Ashcroft invariants,
supporting the interactive application of the method to real-life
architectures (controllers, autonomous cyber-physical systems, etc.)

%% file: philosophers.pdf_t
\begin{picture}(0,0)%
\includegraphics{philosophers.pdf}%
\end{picture}%
\setlength{\unitlength}{2368sp}%
\begingroup\makeatletter\ifx\SetFigFont\undefined%
\gdef\SetFigFont#1#2#3#4#5{%
  \reset@font\fontsize{#1}{#2pt}%
  \fontfamily{#3}\fontseries{#4}\fontshape{#5}%
  \selectfont}%
\fi\endgroup%
\begin{picture}(4449,2949)(889,-2548)
\put(4276, 89){\makebox(0,0)[b]{\smash{{\SetFigFont{7}{8.4}{\rmdefault}{\mddefault}{\updefault}{\color[rgb]{0,0,0}$\ell(\succ(k))$}%
}}}}
\put(3076,-436){\makebox(0,0)[b]{\smash{{\SetFigFont{7}{8.4}{\rmdefault}{\mddefault}{\updefault}{\color[rgb]{0,0,0}$w$}%
}}}}
\put(3076,-1636){\makebox(0,0)[b]{\smash{{\SetFigFont{7}{8.4}{\rmdefault}{\mddefault}{\updefault}{\color[rgb]{0,0,0}$e$}%
}}}}
\put(3451,-1036){\makebox(0,0)[b]{\smash{{\SetFigFont{7}{8.4}{\rmdefault}{\mddefault}{\updefault}{\color[rgb]{0,0,0}$g$}%
}}}}
\put(2701,-1036){\makebox(0,0)[b]{\smash{{\SetFigFont{7}{8.4}{\rmdefault}{\mddefault}{\updefault}{\color[rgb]{0,0,0}$p$}%
}}}}
\put(2776, 89){\makebox(0,0)[b]{\smash{{\SetFigFont{7}{8.4}{\rmdefault}{\mddefault}{\updefault}{\color[rgb]{0,0,0}$p(k)$}%
}}}}
\put(3376, 89){\makebox(0,0)[b]{\smash{{\SetFigFont{7}{8.4}{\rmdefault}{\mddefault}{\updefault}{\color[rgb]{0,0,0}$g(k)$}%
}}}}
\put(4726,-436){\makebox(0,0)[b]{\smash{{\SetFigFont{7}{8.4}{\rmdefault}{\mddefault}{\updefault}{\color[rgb]{0,0,0}$f$}%
}}}}
\put(4726,-1636){\makebox(0,0)[b]{\smash{{\SetFigFont{7}{8.4}{\rmdefault}{\mddefault}{\updefault}{\color[rgb]{0,0,0}$b$}%
}}}}
\put(5101,-1036){\makebox(0,0)[b]{\smash{{\SetFigFont{7}{8.4}{\rmdefault}{\mddefault}{\updefault}{\color[rgb]{0,0,0}$t$}%
}}}}
\put(4351,-1036){\makebox(0,0)[b]{\smash{{\SetFigFont{7}{8.4}{\rmdefault}{\mddefault}{\updefault}{\color[rgb]{0,0,0}$\ell$}%
}}}}
\put(3076,-2086){\makebox(0,0)[b]{\smash{{\SetFigFont{7}{8.4}{\rmdefault}{\mddefault}{\updefault}{\color[rgb]{0,0,0}$\mathsf{Philosopher}(k)$}%
}}}}
\put(1501,-2086){\makebox(0,0)[b]{\smash{{\SetFigFont{7}{8.4}{\rmdefault}{\mddefault}{\updefault}{\color[rgb]{0,0,0}$\mathsf{Fork}(k)$}%
}}}}
\put(4801,-2086){\makebox(0,0)[b]{\smash{{\SetFigFont{7}{8.4}{\rmdefault}{\mddefault}{\updefault}{\color[rgb]{0,0,0}$\mathsf{Fork}((k+1)\mod N)$}%
}}}}
\put(3301,-2461){\makebox(0,0)[b]{\smash{{\SetFigFont{8}{9.6}{\rmdefault}{\mddefault}{\updefault}{\color[rgb]{0,0,0}$\interform_{\mathit{philo}} = \exists i ~.~ [g(i) \wedge t(i)\wedge t(\succ(i))] \vee [p(i) \wedge \ell(i) \wedge \ell(\succ(i))]$}%
}}}}
\put(1501,-436){\makebox(0,0)[b]{\smash{{\SetFigFont{7}{8.4}{\rmdefault}{\mddefault}{\updefault}{\color[rgb]{0,0,0}$f$}%
}}}}
\put(1501,-1636){\makebox(0,0)[b]{\smash{{\SetFigFont{7}{8.4}{\rmdefault}{\mddefault}{\updefault}{\color[rgb]{0,0,0}$b$}%
}}}}
\put(1876,-1036){\makebox(0,0)[b]{\smash{{\SetFigFont{7}{8.4}{\rmdefault}{\mddefault}{\updefault}{\color[rgb]{0,0,0}$t$}%
}}}}
\put(1126,-1036){\makebox(0,0)[b]{\smash{{\SetFigFont{7}{8.4}{\rmdefault}{\mddefault}{\updefault}{\color[rgb]{0,0,0}$\ell$}%
}}}}
\put(1201, 89){\makebox(0,0)[b]{\smash{{\SetFigFont{7}{8.4}{\rmdefault}{\mddefault}{\updefault}{\color[rgb]{0,0,0}$\ell(k)$}%
}}}}
\put(1801, 89){\makebox(0,0)[b]{\smash{{\SetFigFont{7}{8.4}{\rmdefault}{\mddefault}{\updefault}{\color[rgb]{0,0,0}$t(k)$}%
}}}}
\put(5176, 89){\makebox(0,0)[b]{\smash{{\SetFigFont{7}{8.4}{\rmdefault}{\mddefault}{\updefault}{\color[rgb]{0,0,0}$t(\succ(k))$}%
}}}}
\end{picture}%

%% file: philosophers-pn.pdf_t
\begin{picture}(0,0)%
\includegraphics{philosophers-pn.pdf}%
\end{picture}%
\setlength{\unitlength}{2368sp}%
\begingroup\makeatletter\ifx\SetFigFont\undefined%
\gdef\SetFigFont#1#2#3#4#5{%
  \reset@font\fontsize{#1}{#2pt}%
  \fontfamily{#3}\fontseries{#4}\fontshape{#5}%
  \selectfont}%
\fi\endgroup%
\begin{picture}(4533,1681)(961,-1794)
\put(5003,-1617){\makebox(0,0)[b]{\smash{{\SetFigFont{6}{7.2}{\rmdefault}{\mddefault}{\updefault}{\color[rgb]{0,0,0}$(b,\succ(k))$}%
}}}}
\put(4984,-431){\makebox(0,0)[b]{\smash{{\SetFigFont{6}{7.2}{\rmdefault}{\mddefault}{\updefault}{\color[rgb]{0,0,0}$(f,\succ(k))$}%
}}}}
\put(3076,-436){\makebox(0,0)[b]{\smash{{\SetFigFont{6}{7.2}{\rmdefault}{\mddefault}{\updefault}{\color[rgb]{0,0,0}$(w,k)$}%
}}}}
\put(3076,-1636){\makebox(0,0)[b]{\smash{{\SetFigFont{6}{7.2}{\rmdefault}{\mddefault}{\updefault}{\color[rgb]{0,0,0}$(e,k)$}%
}}}}
\put(1501,-1636){\makebox(0,0)[b]{\smash{{\SetFigFont{6}{7.2}{\rmdefault}{\mddefault}{\updefault}{\color[rgb]{0,0,0}$(b,k)$}%
}}}}
\put(976,-1036){\makebox(0,0)[b]{\smash{{\SetFigFont{7}{8.4}{\rmdefault}{\mddefault}{\updefault}{\color[rgb]{0,0,0}$\ldots$}%
}}}}
\put(5176,-1036){\makebox(0,0)[b]{\smash{{\SetFigFont{7}{8.4}{\rmdefault}{\mddefault}{\updefault}{\color[rgb]{0,0,0}$\ldots$}%
}}}}
\put(1501,-436){\makebox(0,0)[b]{\smash{{\SetFigFont{6}{7.2}{\rmdefault}{\mddefault}{\updefault}{\color[rgb]{0,0,0}$(f,k)$}%
}}}}
\put(2246,-1016){\makebox(0,0)[b]{\smash{{\SetFigFont{5}{6.0}{\rmdefault}{\mddefault}{\updefault}{\color[rgb]{0,0,0}$p_k,\ell_k,\ell_{\succ(k)}$}%
}}}}
\put(3991,-1001){\makebox(0,0)[b]{\smash{{\SetFigFont{5}{6.0}{\rmdefault}{\mddefault}{\updefault}{\color[rgb]{0,0,0}$g_k,t_k,t_{\succ(k)}$}%
}}}}
\end{picture}%

%% file: philosophers2.pdf_t
\begin{picture}(0,0)%
\includegraphics{philosophers2.pdf}%
\end{picture}%
\setlength{\unitlength}{2368sp}%
\begingroup\makeatletter\ifx\SetFigFont\undefined%
\gdef\SetFigFont#1#2#3#4#5{%
  \reset@font\fontsize{#1}{#2pt}%
  \fontfamily{#3}\fontseries{#4}\fontshape{#5}%
  \selectfont}%
\fi\endgroup%
\begin{picture}(9325,3933)(889,-3457)
\put(6001,-3361){\makebox(0,0)[b]{\smash{{\SetFigFont{10}{12.0}{\rmdefault}{\mddefault}{\updefault}{\color[rgb]{0,0,0}$\neg\mathit{inf}(x) \wedge [(g\ell(x) \wedge g(x)) \vee (gr(x) \wedge g(s(x))) \vee (p(x) \wedge \ell(x) \wedge \ell(s(x)))]$
}%
}}}}
\put(1501,-436){\makebox(0,0)[b]{\smash{{\SetFigFont{7}{8.4}{\rmdefault}{\mddefault}{\updefault}{\color[rgb]{0,0,0}$f$}%
}}}}
\put(1501,-1636){\makebox(0,0)[b]{\smash{{\SetFigFont{7}{8.4}{\rmdefault}{\mddefault}{\updefault}{\color[rgb]{0,0,0}$b$}%
}}}}
\put(1801,-661){\makebox(0,0)[lb]{\smash{{\SetFigFont{6}{7.2}{\rmdefault}{\mddefault}{\updefault}{\color[rgb]{0,0,0}get}%
}}}}
\put(1276,-1186){\makebox(0,0)[lb]{\smash{{\SetFigFont{6}{7.2}{\rmdefault}{\mddefault}{\updefault}{\color[rgb]{0,0,0}leave}%
}}}}
\put(3076,-436){\makebox(0,0)[b]{\smash{{\SetFigFont{7}{8.4}{\rmdefault}{\mddefault}{\updefault}{\color[rgb]{0,0,0}$w$}%
}}}}
\put(3076,-1636){\makebox(0,0)[b]{\smash{{\SetFigFont{7}{8.4}{\rmdefault}{\mddefault}{\updefault}{\color[rgb]{0,0,0}$e$}%
}}}}
\put(3601,-1036){\makebox(0,0)[b]{\smash{{\SetFigFont{7}{8.4}{\rmdefault}{\mddefault}{\updefault}{\color[rgb]{0,0,0}$h$}%
}}}}
\put(2851,-1186){\makebox(0,0)[lb]{\smash{{\SetFigFont{6}{7.2}{\rmdefault}{\mddefault}{\updefault}{\color[rgb]{0,0,0}put}%
}}}}
\put(3376,-436){\makebox(0,0)[lb]{\smash{{\SetFigFont{6}{7.2}{\rmdefault}{\mddefault}{\updefault}{\color[rgb]{0,0,0}getright}%
}}}}
\put(3376,-1561){\makebox(0,0)[lb]{\smash{{\SetFigFont{6}{7.2}{\rmdefault}{\mddefault}{\updefault}{\color[rgb]{0,0,0}getleft}%
}}}}
\put(9376,-436){\makebox(0,0)[b]{\smash{{\SetFigFont{7}{8.4}{\rmdefault}{\mddefault}{\updefault}{\color[rgb]{0,0,0}$w$}%
}}}}
\put(9376,-1636){\makebox(0,0)[b]{\smash{{\SetFigFont{7}{8.4}{\rmdefault}{\mddefault}{\updefault}{\color[rgb]{0,0,0}$e$}%
}}}}
\put(9901,-1036){\makebox(0,0)[b]{\smash{{\SetFigFont{7}{8.4}{\rmdefault}{\mddefault}{\updefault}{\color[rgb]{0,0,0}$h$}%
}}}}
\put(9151,-1186){\makebox(0,0)[lb]{\smash{{\SetFigFont{6}{7.2}{\rmdefault}{\mddefault}{\updefault}{\color[rgb]{0,0,0}put}%
}}}}
\put(9676,-436){\makebox(0,0)[lb]{\smash{{\SetFigFont{6}{7.2}{\rmdefault}{\mddefault}{\updefault}{\color[rgb]{0,0,0}getleft}%
}}}}
\put(9676,-1561){\makebox(0,0)[lb]{\smash{{\SetFigFont{6}{7.2}{\rmdefault}{\mddefault}{\updefault}{\color[rgb]{0,0,0}getright}%
}}}}
\put(1666,-11){\makebox(0,0)[b]{\smash{{\SetFigFont{6}{7.2}{\rmdefault}{\mddefault}{\updefault}{\color[rgb]{0,0,0}$g(0)$}%
}}}}
\put(3166,-11){\makebox(0,0)[b]{\smash{{\SetFigFont{6}{7.2}{\rmdefault}{\mddefault}{\updefault}{\color[rgb]{0,0,0}$\overline{gr}(0)$}%
}}}}
\put(3616,-11){\makebox(0,0)[b]{\smash{{\SetFigFont{6}{7.2}{\rmdefault}{\mddefault}{\updefault}{\color[rgb]{0,0,0}$\overline{g\ell}(0)$}%
}}}}
\put(9466,-11){\makebox(0,0)[b]{\smash{{\SetFigFont{6}{7.2}{\rmdefault}{\mddefault}{\updefault}{\color[rgb]{0,0,0}$gr(2)$}%
}}}}
\put(9916,-11){\makebox(0,0)[b]{\smash{{\SetFigFont{6}{7.2}{\rmdefault}{\mddefault}{\updefault}{\color[rgb]{0,0,0}$g\ell(2)$}%
}}}}
\put(9376,-2536){\makebox(0,0)[b]{\smash{{\SetFigFont{7}{8.4}{\rmdefault}{\mddefault}{\updefault}{\color[rgb]{0,0,0}$\mathsf{Philosopher}_{lr}(2)$}%
}}}}
\put(1501,-2536){\makebox(0,0)[b]{\smash{{\SetFigFont{7}{8.4}{\rmdefault}{\mddefault}{\updefault}{\color[rgb]{0,0,0}$\mathsf{Fork}(0)$}%
}}}}
\put(4801,-2536){\makebox(0,0)[b]{\smash{{\SetFigFont{7}{8.4}{\rmdefault}{\mddefault}{\updefault}{\color[rgb]{0,0,0}$\mathsf{Fork}(1)$}%
}}}}
\put(6226,-2536){\makebox(0,0)[b]{\smash{{\SetFigFont{7}{8.4}{\rmdefault}{\mddefault}{\updefault}{\color[rgb]{0,0,0}$\mathsf{Philosopher}_{lr}(1)$}%
}}}}
\put(3076,-2536){\makebox(0,0)[b]{\smash{{\SetFigFont{7}{8.4}{\rmdefault}{\mddefault}{\updefault}{\color[rgb]{0,0,0}$\mathsf{Philosopher}_{rl}(0)$}%
}}}}
\put(7876,-2536){\makebox(0,0)[b]{\smash{{\SetFigFont{7}{8.4}{\rmdefault}{\mddefault}{\updefault}{\color[rgb]{0,0,0}$\mathsf{Fork}(2)$}%
}}}}
\put(1426,-2011){\makebox(0,0)[b]{\smash{{\SetFigFont{6}{7.2}{\rmdefault}{\mddefault}{\updefault}{\color[rgb]{0,0,0}$\ell(0)$}%
}}}}
\put(2926,-2011){\makebox(0,0)[b]{\smash{{\SetFigFont{6}{7.2}{\rmdefault}{\mddefault}{\updefault}{\color[rgb]{0,0,0}$\overline{p}(0)$}%
}}}}
\put(9226,-2011){\makebox(0,0)[b]{\smash{{\SetFigFont{6}{7.2}{\rmdefault}{\mddefault}{\updefault}{\color[rgb]{0,0,0}$p(2)$}%
}}}}
\put(5401,-3061){\makebox(0,0)[b]{\smash{{\SetFigFont{10}{12.0}{\rmdefault}{\mddefault}{\updefault}{\color[rgb]{0,0,0}$\interform_{\mathit{philo}}^{\mathit{alt}} = \exists x ~.~ \mathit{inf}(x) \wedge [(\overline{g\ell}(x) \wedge g(x)) \vee (\overline{gr}(x) \wedge g(s(x))) \vee (\overline{p}(x) \wedge \ell(x) \wedge \ell(s(x)))] ~\vee$
}%
}}}}
\put(7801,-436){\makebox(0,0)[b]{\smash{{\SetFigFont{7}{8.4}{\rmdefault}{\mddefault}{\updefault}{\color[rgb]{0,0,0}$f$}%
}}}}
\put(7801,-1636){\makebox(0,0)[b]{\smash{{\SetFigFont{7}{8.4}{\rmdefault}{\mddefault}{\updefault}{\color[rgb]{0,0,0}$b$}%
}}}}
\put(8101,-661){\makebox(0,0)[lb]{\smash{{\SetFigFont{6}{7.2}{\rmdefault}{\mddefault}{\updefault}{\color[rgb]{0,0,0}get}%
}}}}
\put(7576,-1186){\makebox(0,0)[lb]{\smash{{\SetFigFont{6}{7.2}{\rmdefault}{\mddefault}{\updefault}{\color[rgb]{0,0,0}leave}%
}}}}
\put(4726,-436){\makebox(0,0)[b]{\smash{{\SetFigFont{7}{8.4}{\rmdefault}{\mddefault}{\updefault}{\color[rgb]{0,0,0}$f$}%
}}}}
\put(4726,-1636){\makebox(0,0)[b]{\smash{{\SetFigFont{7}{8.4}{\rmdefault}{\mddefault}{\updefault}{\color[rgb]{0,0,0}$b$}%
}}}}
\put(5026,-661){\makebox(0,0)[lb]{\smash{{\SetFigFont{6}{7.2}{\rmdefault}{\mddefault}{\updefault}{\color[rgb]{0,0,0}get}%
}}}}
\put(4501,-1186){\makebox(0,0)[lb]{\smash{{\SetFigFont{6}{7.2}{\rmdefault}{\mddefault}{\updefault}{\color[rgb]{0,0,0}leave}%
}}}}
\put(6151,-436){\makebox(0,0)[b]{\smash{{\SetFigFont{7}{8.4}{\rmdefault}{\mddefault}{\updefault}{\color[rgb]{0,0,0}$w$}%
}}}}
\put(6151,-1636){\makebox(0,0)[b]{\smash{{\SetFigFont{7}{8.4}{\rmdefault}{\mddefault}{\updefault}{\color[rgb]{0,0,0}$e$}%
}}}}
\put(6676,-1036){\makebox(0,0)[b]{\smash{{\SetFigFont{7}{8.4}{\rmdefault}{\mddefault}{\updefault}{\color[rgb]{0,0,0}$h$}%
}}}}
\put(5926,-1186){\makebox(0,0)[lb]{\smash{{\SetFigFont{6}{7.2}{\rmdefault}{\mddefault}{\updefault}{\color[rgb]{0,0,0}put}%
}}}}
\put(6451,-436){\makebox(0,0)[lb]{\smash{{\SetFigFont{6}{7.2}{\rmdefault}{\mddefault}{\updefault}{\color[rgb]{0,0,0}getleft}%
}}}}
\put(6451,-1561){\makebox(0,0)[lb]{\smash{{\SetFigFont{6}{7.2}{\rmdefault}{\mddefault}{\updefault}{\color[rgb]{0,0,0}getright}%
}}}}
\put(4891,-11){\makebox(0,0)[b]{\smash{{\SetFigFont{6}{7.2}{\rmdefault}{\mddefault}{\updefault}{\color[rgb]{0,0,0}$g(1)$}%
}}}}
\put(6241,-11){\makebox(0,0)[b]{\smash{{\SetFigFont{6}{7.2}{\rmdefault}{\mddefault}{\updefault}{\color[rgb]{0,0,0}$gr(1)$}%
}}}}
\put(6691,-11){\makebox(0,0)[b]{\smash{{\SetFigFont{6}{7.2}{\rmdefault}{\mddefault}{\updefault}{\color[rgb]{0,0,0}$g\ell(1)$}%
}}}}
\put(7966,-11){\makebox(0,0)[b]{\smash{{\SetFigFont{6}{7.2}{\rmdefault}{\mddefault}{\updefault}{\color[rgb]{0,0,0}$g(2)$}%
}}}}
\put(4726,-2011){\makebox(0,0)[b]{\smash{{\SetFigFont{6}{7.2}{\rmdefault}{\mddefault}{\updefault}{\color[rgb]{0,0,0}$\ell(1)$}%
}}}}
\put(6076,-2011){\makebox(0,0)[b]{\smash{{\SetFigFont{6}{7.2}{\rmdefault}{\mddefault}{\updefault}{\color[rgb]{0,0,0}$p(1)$}%
}}}}
\put(7876,-2011){\makebox(0,0)[b]{\smash{{\SetFigFont{6}{7.2}{\rmdefault}{\mddefault}{\updefault}{\color[rgb]{0,0,0}$\ell(2)$}%
}}}}
\end{picture}%

%% file: window.pdf_t
\begin{picture}(0,0)%
\includegraphics{window.pdf}%
\end{picture}%
\setlength{\unitlength}{2368sp}%
\begingroup\makeatletter\ifx\SetFigFont\undefined%
\gdef\SetFigFont#1#2#3#4#5{%
  \reset@font\fontsize{#1}{#2pt}%
  \fontfamily{#3}\fontseries{#4}\fontshape{#5}%
  \selectfont}%
\fi\endgroup%
\begin{picture}(4144,2071)(961,-1818)
\put(2686,-1740){\makebox(0,0)[b]{\smash{{\SetFigFont{7}{8.4}{\rmdefault}{\mddefault}{\updefault}{\color[rgb]{0,0,0}$\psi \isdef \exists \zeta ~.~ \mathit{inf}(\zeta) \wedge \zeta < c_1 \wedge c_1 < c_2 \wedge c_2 = c_3 \wedge c_2 = s(c_1)$}%
}}}}
\put(2851,-136){\makebox(0,0)[b]{\smash{{\SetFigFont{7}{8.4}{\rmdefault}{\mddefault}{\updefault}{\color[rgb]{0,0,0}$f(c_2)$}%
}}}}
\put(2851,-1336){\makebox(0,0)[b]{\smash{{\SetFigFont{7}{8.4}{\rmdefault}{\mddefault}{\updefault}{\color[rgb]{0,0,0}$b(c_2)$}%
}}}}
\put(4276,-136){\makebox(0,0)[b]{\smash{{\SetFigFont{7}{8.4}{\rmdefault}{\mddefault}{\updefault}{\color[rgb]{0,0,0}$w(c_3)$}%
}}}}
\put(4801,-736){\makebox(0,0)[b]{\smash{{\SetFigFont{7}{8.4}{\rmdefault}{\mddefault}{\updefault}{\color[rgb]{0,0,0}$h(c_3)$}%
}}}}
\put(1201,-1336){\makebox(0,0)[b]{\smash{{\SetFigFont{7}{8.4}{\rmdefault}{\mddefault}{\updefault}{\color[rgb]{0,0,0}$e(c_1)$}%
}}}}
\put(1726,-736){\makebox(0,0)[b]{\smash{{\SetFigFont{7}{8.4}{\rmdefault}{\mddefault}{\updefault}{\color[rgb]{0,0,0}$h(c_1)$}%
}}}}
\put(1224,-121){\makebox(0,0)[b]{\smash{{\SetFigFont{7}{8.4}{\rmdefault}{\mddefault}{\updefault}{\color[rgb]{0,0,0}$w(c_1)$}%
}}}}
\put(4276,-1336){\makebox(0,0)[b]{\smash{{\SetFigFont{7}{8.4}{\rmdefault}{\mddefault}{\updefault}{\color[rgb]{0,0,0}$e(c_3)$}%
}}}}
\end{picture}%